\documentclass[letterpaper,11pt]{article}
\usepackage[DIV12]{typearea}
\usepackage{longtable}
\usepackage{booktabs,colortbl}
\usepackage{multirow}
\usepackage{amsmath}
\usepackage{amssymb}
\usepackage{bbm}
\usepackage[utf8]{inputenc}
\usepackage{pdflscape}
\usepackage{xspace}
\usepackage[caption=false]{subfig}
\usepackage{nicefrac}
\usepackage{graphicx}
\usepackage{cite}  
\usepackage[small]{caption2}
\usepackage{mathtools}
\usepackage{nccfoots}
\usepackage{tikz}
\usetikzlibrary{calc,positioning}

\newcommand{\dd}{\mathrm{d}}

\usepackage[all,cmtip]{xy}
\newlength{\xywd}
\newcommand{\xyrightarrow}[2][]{%
  \sbox{0}{$\scriptstyle#1$}%
  \xywd=\wd0
  \sbox{0}{$\scriptstyle#2$}%
  \ifdim\wd0>\xywd \xywd=\wd0 \fi
  \xymatrix@C\dimexpr\xywd+1em\relax{{}\ar[r]^{#2}_{#1}&{}}%
}

\renewcommand{\thefootnote}{\alph{footnote}}

\newcommand{\SU}[1]{\ensuremath{\mathrm{SU}(#1)}}

\newcommand{\x}{\ensuremath{\times}}

\usepackage{xcolor}
\definecolor{darkgreen}{HTML}{109930}
\definecolor{pink}{rgb}{0.858, 0.188, 0.478}

\advance \headheight by 3.0truept       
\usepackage[pdftex,hidelinks,colorlinks=true,citecolor=darkgreen]{hyperref}
\hypersetup{
    pdftitle = {Spectral distortions from promising single and multifield inflationary models},
    pdfauthor = {Baur, Garcia, Henriquez-Ortiz, Hernandez-Neri, Ramos-Sanchez}
}

\usepackage[toc,page]{appendix}    

\usepackage{cleveref}

\begin{document}

\begin{titlepage}
	
\vspace*{-3.0cm}
\begin{flushright}
	TUM-HEP 1476/23
\end{flushright}

\vspace*{1.0cm}

\begin{center}
{\Large\textbf{\boldmath Spectral distortions\\[1mm] from promising single and multifield inflationary models\unboldmath}}

\vspace{1cm}
\textbf{Alexander Baur$^{\dagger,\S}$\footnote{\texttt{alexander.baur@tum.de}}},
\textbf{Marcos A.~G. Garc\'ia$^\dagger$\footnote{\texttt{marcos.garcia@fisica.unam.mx}}},
\textbf{Ra\'ul Henr\'iquez--Ortiz$^*$\footnote{\texttt{raul.henriquez@ues.edu.sv}}},\\[1mm]
\textbf{Mauricio Hern\'andez--Neri$^\dagger$\footnote{\texttt{mauriciohn@estudiantes.fisica.unam.mx}}}
and
\textbf{Sa\'ul Ramos--S\'anchez$^\dagger$\footnote{\texttt{ramos@fisica.unam.mx}}}\\[5mm]

\textit{\small $^\dagger$ Instituto de F\'isica, Universidad Nacional Aut\'onoma de M\'exico,\\ Cd.~de M\'exico C.P.~04510, M\'exico}\\[1mm]
\textit{\small $^\S$ Physik Department, Technische Universit\"at M\"unchen,\\ James-Franck-Stra\ss e 1, 85748 Garching, Germany}\\[1mm]
\textit{\small $^*$ Escuela de F\'isica, Facultad de Ciencias Naturales y Matem\'atica, Universidad de El Salvador,\\
final de Av. M\'artires y H\'eroes del 30 julio, San Salvador, C.P.~1101, El Salvador}
\end{center}

\vspace{1cm}

\vspace*{1.0cm}

\begin{abstract}
Forthcoming missions probing the absolute intensity of the CMB are expected to be able to measure 
spectral distortions, which are deviations from its blackbody distribution. As cosmic inflation can induce spectral 
distortions, these experiments offer a possibility to further test the various promising inflationary 
proposals, whose predictions need to be carefully determined. After numerically fitting all inflationary 
observables to match current observations, we compute the predicted spectral distortions of 
various promising single and multifield inflationary models. The predictions of single-field 
inflationary models display deviations between 0.5\% and 20\% with respect to the standard cosmological 
model in the observable window, where multi-natural and axion-monodromy inflation stand out in this respect.
In the case of multifield inflation, we observe a richer structure of the power spectrum, which, in 
the case of so-called hybrid attractors, yields spectral distortions about 100 times more intense than
the standard signal. These observations open up questions about the relation among our results
and other cosmological observables that are also to be probed soon, such as the production of
primordial black holes and gravitational waves.

\end{abstract}

\end{titlepage}

\newpage
\setcounter{footnote}{0} 
\renewcommand{\thefootnote}{\arabic{footnote}}

\section{Introduction}

The Cosmic Microwave Background (CMB) is known to display roughly a blackbody-radiation distribution
due to the isotropicity and thermal equilibrium of the photon-matter fluid in the early Universe 
induced by particle physics processes, such as Bremstrahlung (BR), Compton scattering (CS) 
and double Compton scattering (DC). However, small departures from this distribution, known as 
spectral distortions (SDs), could have been sowed by early out-of-equilibrium energy injections 
into the photon-matter fluid~\cite{Chluba:2019kpb}. After the early efforts to detect them by 
COBE/FIRAS~\cite{Fixsen:1996nj,Bianchini:2022dqh}, SDs might soon be detected by forthcoming 
space missions, such as PRISM~\cite{PRISM:2013ybg}, PIXIE~\cite{Kogut:2011xw},
its enhanced version Super-PIXIE and the ESA Voyage 2050 program, which may reach 
a sensitivity of up to seven orders of magnitude better than the 
accuracy of FIRAS~\cite{Chluba:2019nxa,Fu:2020wkq}. 

Processes that inject energy into the matter-photon fluid and affect it differently at 
different scales can produce SDs due to 
various standard~\cite{Hu:1994bz,Khatri:2012tv,Chluba:2013wsa,Chluba:2013dna,Rubino-Martin:2006hng} 
and non-standard~\cite{Diacoumis:2017hff,Chluba:2019nxa,Chluba:2013pya,Lucca:2019rxf} mechanisms. 
SDs are sensitive to the structure of the primordial power spectrum $P_\mathcal{R}(k)$ by 
Silk damping~\cite{Silk:1967kq}, which damps acoustic waves that enter and are smaller than the 
sound horizon, such as those induced by any small perturbation. After this energy is thermally 
redistributed, the radiation spectrum becomes a mixture of blackbody radiation from regions with 
various temperatures, which is no longer blackbody radiation and can thus be regarded as SDs.
Hence, the perturbations contained in $\mathcal{P}_\mathcal{R}(k)$ are translated into 
potentially observable SDs.

In standard cosmology, the shape and intensity of the primordial power spectrum~\cite{Kosowsky:1995aa}
are defined by the features of inflation~\cite{Guth:1980zm,Linde:1981mu,Albrecht:1982wi} 
(see also~\cite{Liddle:1999mq,Tsujikawa:2003jp,Vazquez:2018qdg,Achucarro:2022qrl}). 
Despite the great precision already achieved by CMB observatories, such as {\it Planck}~\cite{Planck:2018jri} 
and BICEP/Keck~\cite{BICEP:2021xfz}, there is still
a plethora of inflationary models~\cite{Martin:2013tda,Martin:2013nzq} 
that successfully deliver the measured values of the (inflationary) parameters of $\mathcal P_\mathcal{R}(k)$.
These include the amplitude $A_{S_\star}$, the tilt or spectral index $n_s$ and its running 
$\frac{\mathrm{d} n_s}{\mathrm{d}\,\mathrm{ln}\,k}$, and the ratio $r$ of the tensor 
and scalar amplitudes. The challenge of discriminating among such a variety of possible 
models could receive important input from probing SDs. Hence, it becomes imperative 
to work out the predictions for the SDs of the inflationary scenarios that best describe all 
CMB measurements so far. This endeavor has already been started (see 
e.g.~\cite{Bae:2017tll,Henriquez-Ortiz:2022ulz,Zegeye:2021yml,Clesse:2014pna}). For example,
using the large class of inflationary models with up to three free parameters subject to early 
CMB constraints of~\cite{Martin:2013tda}, predictions for SDs and their potential detection
by PIXIE were studied in~\cite{Clesse:2014pna}. While our present study has a similar philosophy,
in this work we investigate various models not considered previously, and we take into account
the updated CMB constraints~\cite{Planck:2018jri,BICEP:2021xfz} and forecasts for SDs by both PIXIE and
its enhanced versions with higher sensitivity~\cite{Kogut:2011xw,Chluba:2019nxa,Fu:2020wkq}. 
As discussed in detail in~\cite{Henriquez-Ortiz:2022ulz}, these considerations might open an opportunity 
window to discriminate and falsify similar SDs signals arising from different models.

Determining the spectral distortions of successful inflationary models consists of
two steps. First, assuming that inflation is driven by the {\it slowly rolling} 
dynamics of an inflaton field $\phi$, encoded in its scalar potential $V(\phi)$,
one must establish under what circumstances the model can reproduce the
observations on the inflationary parameters, i.e.\ it is necessary to 
numerically fit the parameters of $V(\phi)$ that lead to the observables within 
the accuracy of {\it Planck}-BICEP/Keck data~\cite{BICEP:2021xfz}. One can then 
solve the Mukhanov-Sasaki equation with the right parameters to obtain the corresponding
primordial power spectrum, and finally use this to compute the contribution 
$\Delta I$ to the photon intensity produced by the SDs predicted by the model. 
To do so, one can either rely on approximations for each contribution to SDs, 
as in~\cite{Chluba:2012we,Cho:2017zkj,Schoneberg:2020nyg,Henriquez-Ortiz:2022ulz}, or numerically compute them by using
the \texttt{CLASS} code~\cite{CLASS,Blas:2011rf,Lucca:2019rxf} to arrive at more 
accurate results. We shall prefer the latter.

Our paper is organized as follows. In order to fix our notation, in Section~\ref{sec:distortions}
we present a discussion of basic details of SDs. Then, in Section~\ref{sec:SingleFieldInflation},
we proceed to study a selection of known single-field models of cosmic inflation, 
addressing first the question of their compatibility with known CMB observations
and then showing the predicted SDs. Our study would not be complete if we did not
address the outcome from some multifield inflationary models, which we do in 
Section~\ref{sec:multifield}, where encouraging SDs signals are uncovered. To stress 
our main results, we summarize them in Section~\ref{sec:results}.

\section{Spectral distortions}
\label{sec:distortions}
The dissipation of acoustic waves with adiabatic initial conditions generates SDs~\cite{Chluba:2012we,Dimastrogiovanni:2016aul,Daly:1991,Chluba:2016bvg,Chluba:2019nxa,Cho:2017zkj}. 
We can determine the full photon intensity spectrum $I(z,x)$ of the CMB in terms of the 
blackbody distribution $I_0$ and the total contribution of SDs $\Delta I(z,x)$, i.e.
\begin{equation}\label{eq:sdn1}
I(z,x)= I_{0}(x)+\Delta I(z,x) \,,
\end{equation}
where $z$ denotes the redshift and $I_{0}(x):=\frac{2h\nu^{3}}{c^{2}} \frac{1}{e^x - 1}$ with the dimensionless frequency
$x=h\nu/k_\textrm{B}T$. The distortion of the photon intensity spectrum $\Delta I(z,x)$
is given at first order in terms of the temperature shift $\Delta T$ (which does not modify the 
blackbody distribution), and the contributions to SDs $y$ and $\mu$, associated with the 
various processes and epochs that deviate the CMB spectrum from its blackbody structure, see 
ref.~\cite{Lucca:2019rxf}. Hence, the contribution of SDs to the CMB spectrum is modeled solving the full evolution of the photon phase-space distribution in the Boltzmann equation. Since the distortion is a small correction, the Boltzmann equation may be linearized around the blackbody solution and integrated using a Green's function
\begin{equation}
\label{eq:sdn2}
\Delta I(z,x)   ~=~ \Delta I_{\rm R}(z,x) +  \int_{z}^{\infty} \dd z' G^{\star}_{\rm th}(x,z')\frac{\nicefrac{\dd Q(z')}{\dd z'}}{\rho_{\gamma(z')}}\,.
\end{equation}
Here $\rho_{\gamma}^{-1} \dd Q(z')/ \dd z'$ is the effective heating rate, $G^{\star}_{\rm th}(x,z')$ is the kernel in the integral, which includes all physics of the thermalization process in the Universe evolution, while $\Delta I_{\rm R}$ is the remainder from the linearization. The integral in Eq.~\eqref{eq:sdn2} can be parametrized in term of $\Delta T$, $y$, and $\mu$ as
\begin{equation}\label{eq:sd1n}
  \Delta I(z,x) - \Delta I_{\rm R}(z,x) ~=~
 \Delta I_{y}(z,x) + \Delta I_{\mu}(z,x) + \Delta I_{\rm T}(z,x)\,.
\end{equation}
The function $G^{\star}_{\rm th}(x,z')$ has been calculated in \cite{Chluba:2013vsa}, as a function of the redshift
\begin{equation}
\label{eq:sd3}
G^{\star}_{\rm th}(x,z') ~=~\frac{ G_{T}(x)}{4} \mathcal{J}_{T}(z')+\frac{Y_{SZ}(x)}{4} \mathcal{J}_{y}(z')+\alpha M(x) \mathcal{J}_{\mu}(z')\,,
\end{equation}
where the functions $ G_{T}(x)$, $Y_{SZ}(x)$ and $M(x)$ are given by
\begin{subequations}\label{eq:sd3nn}
\begin{eqnarray}
  G_{T}(x) &=& \frac{2 h\nu^{3}}{c^{2}} \frac{x e^{x}}{(e^{x}-1)^{2}}  \,, \\
  Y_{SZ}(x) &=&   G_{T}(x) \left( x \operatorname{coth} \left(  \frac{x}{2} \right) -4 \right)   \,, \\
  M(x) &=&  \frac{G_{T}(x)}{x} \left(\frac{x}{\beta} - 1 \right) \,,
\end{eqnarray}
\end{subequations}
with $\alpha \approx 1.401$ and $\beta \approx 2.192$. The parametrization of $G^{\star}_{\rm th}(x,z')$ allows the computation of $\Delta I_{\rm R}$, the corrective term associated with the residual SDs, when the full distortion is determined by means of numerical codes ~\cite{Chluba:2013pya,Chluba:2013vsa,Chluba:2015hma}.
The transition between $\mu$ and $y$ distortions is gradual (see \cite{Chluba:2013pya} for details). This transitional region in $z$ is represented by the weighting functions $\mathcal{J}_{a}$ (with $a\in\left\lbrace T, y, \mu \right\rbrace $) defined as~\cite{Chluba:2013pya,Lucca:2019rxf}
\begin{subequations}\label{eq:sd2}
\begin{eqnarray}
  \mathcal{J}_{y}(z) &=& \left[1+\left(\frac{1+z}{6\times 10^{4}} \right)^{2.58}\right]^{-1}  \,, \\
  \mathcal{J}_{\mu}(z) &=& \left\lbrace  1-\exp{\left[ -\left(\frac{1+z}{5.8\times 10^{4}} \right)^{1.88} \right]}  \right\rbrace f(z)  \,, \\
  \mathcal{J}_{T}(z) &=& 1-f(z) \,,
\end{eqnarray}
\end{subequations}
where $f(z) \approx \exp\{-\left(z/z_\mathrm{th}\right)^{\nicefrac{5}{2}}\}$ with $z_\mathrm{th}\approx 1.98 \times 10^{6}$. As mentioned above, the approximation in Eq.~\eqref{eq:sdn2} with Eq.~\eqref{eq:sd3} is  only valid if the distortions are very small (i.e.~$\Delta I/I \ll 1$). When the primordial density perturbations re-enter the horizon, pressure gradients produce pressure waves. The propagation of these waves is affected by dissipation, which causes damping at small scales and creates SDs. This dissipation of primordial acoustic modes influences importantly the evolution of the radiation field for some initial conditions~\cite{Chluba:2012gq,Chluba:2013dna}. 

Finally, we can model the effective heating rate for the energy release arising from the damping of adiabatic modes with the approximation (see~\cite{Fu:2020wkq} for details)
\begin{equation}\label{eq:sd4}
\frac{1}{\rho_{\gamma}}\frac{\dd  Q}{\dd z} ~=~ 4 A^{2}  \int_{k_\mathrm{min}}^{\infty} \frac{k^{4} \dd k}{2 \pi^{2}} \mathcal P_{\mathcal{R}}(k) \left[\partial_z k^{-2}_{D}\right]  e^{-2k^{2}/k^2_D} \,.
\end{equation}
Here, $\mathcal P_{\mathcal{R}}(k)$ is the curvature power spectrum, which encodes the information of the 
primordial scalar fluctuations from inflation. In the case of adiabatic modes, $A\approx(1+\nicefrac{4R_\nu}{15})^{-1}$ 
is a normalization coefficient with $\nicefrac{4R_\nu}{15}$ accounting for the correction due to the anisotropic stress in the neutrino fluid, $k_D \approx 4.048 \times10^{-6}(1+z)^{\nicefrac32}$ ${\rm Mpc}^{-1}$ is the photon damping scale and $k_\mathrm{min} =  1~\text{Mpc}^{-1}$. The dependence on the primordial power spectrum allows us to directly characterize the SDs signal associated with different inflationary models. To avoid the introduction of systematic uncertainties associated with $\Delta I_{\rm R}$, especially in the cases with large distortions, we numerically determine the full profile of the SDs signal with the help of the \texttt{CLASS} code\footnote{Our focus in this work is on the SDs associated with processes such as adiabatic cooling through electron scattering and the dissipation of acoustic waves arising from inflationary models, which leave their imprint at all frequencies. A completion of this study should also include contributions of foregrounds to these distortions, as modeled e.g.~in~\cite{Abitbol:2017vwa}, where it is noticed that these foregrounds become more relevant for frequencies below 30 GHz or so. We expect hence our results to be more interesting above such frequencies. The detailed study of foregrounds and many other contributions to the full absolute intensity of CMB is beyond the scope of the present work.}~\cite{CLASS,Blas:2011rf,Lucca:2019rxf}.

\section{Models of single-field inflation}
\label{sec:SingleFieldInflation}
We begin our characterization of CMB constraints and SDs focusing on a 
collection of single field models. For all of them, the dynamics 
can be characterized by the action
\begin{equation}
S \;=\; \int \mathrm d^4x\, \sqrt{-g}\left[ \frac{R}{2} +  \frac{1}{2}\partial_{\mu}\phi\partial^{\mu}\phi - V(\phi)\right]\,,
\end{equation}
where $g$ denotes the space-time metric, $R$ is the Ricci scalar, and $\phi$ 
is the scalar or pseudo-scalar field that drives inflation, which will be assumed 
hereafter to be homogeneous at leading order. For simplicity we work in units where the reduced Planck mass $M_{\rm P}=1/\sqrt{8\pi G}=1$. At the 
background level, $g$ is assumed to have a flat Friedmann-Robertson-Walker form with scale factor 
$a(t)$. The background (homogeneous) equations of motion correspond to the Klein-Gordon-Friedmann 
system of the form
\begin{subequations}
\label{eqs:KGF}
\begin{align}
\label{eq:KG}
\ddot{\phi} + 3H\dot{\phi} + V_{\phi} \;&=\; 0\,,\\
\label{eq:Friedmann}
\frac{1}{2}\dot{\phi}^2 + V(\phi) \;&=\; 3H^2\,,
\end{align}
\end{subequations}
where the subindex of $V_\phi$ denotes the derivative of $V$ w.r.t.\ $\phi$, and
$H:=\dot{a}/a$ is the Hubble parameter. In the slow-roll approximation, $\epsilon, |\eta|\ll 1$
with
\begin{equation}
\label{eq:slowroll0}
\epsilon ~:=~ \frac{1}{2}\left(\frac{V_{\phi}}{V}\right)^2\qquad\text{and}\qquad
\eta ~:=~ \frac{V_{\phi\phi}}{V}\,,
\end{equation}
the system~\eqref{eqs:KGF} can be approximately integrated analytically. 
For later convenience, we also define the third slow-roll parameter
\begin{equation}
\label{eq:xi}
\xi^2 ~=~ \dfrac{V_\phi\,V_{\phi\phi\phi}}{V^2}\,.
\end{equation}

The number of $e$-folds of expansion between the horizon crossing of the pivot scale $k_\star$, 
at the field value $\phi_\star$, and the end of inflation, with field value $\phi_{\rm end}$, 
can be estimated as
\begin{equation}\label{eq:Nstarslowroll}
N_\star \;\simeq\; \int_{\phi_{\rm end}}^{\phi_\star} \frac{\mathrm d\phi}{\sqrt{2\epsilon}}\,.
\end{equation}
The number of $e$-folds are in turn related to the post inflationary dynamics by the well-known 
relation~\cite{Liddle:2003as,Martin:2010kz}
\begin{align}\label{eq:Nstar} \notag
N_\star ~=~ &\ln\left[\frac{1}{\sqrt{3}}\left(\frac{\pi^2}{30}\right)^{1/4}\left(\frac{43}{11}\right)^{1/3}\frac{T_0}{H_0}\right]-\ln\left(\frac{k_\star}{a_0H_0}\right) - \frac{1}{12}\ln g_{\rm reh}\\
&\qquad  + \frac{1}{4}\ln\left(\frac{V_\star^2}{\rho_{\rm end}}\right) + \frac{1-3w_{\rm int}}{12(1+w_{\rm int})}\ln\left(\frac{\rho_{\rm rad}}{\rho_{\rm end}}\right)\,.
\end{align}
Here, $V_\star:=V(\phi_\star)$. Further, $H_0=67.36\,{\rm km}\,{\rm s}^{-1}{\rm Mpc}^{-1}$~\cite{Planck:2018vyg}, 
$T_0=2.7255\,{\rm K}$~\cite{Fixsen:2009ug}, and $a_0=1$ denote respectively the present 
Hubble parameter, photon temperature, and scale factor. The energy density at the end of 
inflation is denoted by $\rho_{\rm end}$, and the energy density at the beginning of the radiation dominated era by $\rho_{\rm rad}$. The effective number of degrees of freedom during reheating is denoted by $g_{\rm reh}$. The $e$-fold averaged equation of state parameter during reheating corresponds to 
\begin{equation}
w_{\rm int} ~:=~ \frac{1}{N_{\rm rad}-N_{\rm end}}\int_{N_{\rm end}}^{N_{\rm rad}} w(N')\,\mathrm dN'\,.
\end{equation}

The scalar perturbation of phenomenological interest corresponds to the curvature fluctuation. 
At the linear order, its Fourier components can be determined by the expression
\begin{equation}
\mathcal{R}_k \;=\; \frac{H}{|\dot{\phi}|}Q_k\,,
\end{equation}
where $Q$ denotes the gauge-invariant Mukhanov-Sasaki variable. Its dynamics are controlled by the equation of motion~\cite{Lalak:2007vi,Ellis:2014opa}
\begin{equation}\label{eq:MSequation}
\ddot{Q}_k + 3H\dot{Q}_k + \left[\frac{k^2}{a^2} + 3\dot{\phi}^2 - \frac{\dot{\phi}^4}{2H^2} + 2\frac{\dot{\phi}V_{\phi}}{H} + V_{\phi\phi}\right]Q_k \;=\; 0\,,
\end{equation}
with Bunch-Davies initial conditions, $Q_{k\gg aH}=e^{ik\tau}/a\sqrt{2k}$, where $\tau$ denotes conformal time. The curvature power spectrum is defined by the relation
\begin{equation}
\langle \mathcal{R}_k \mathcal{R}^*_{k'}\rangle \;=\; \frac{2\pi^2}{k^3}\mathcal{P}_{\mathcal{R}}\delta(k-k')\,.
\end{equation}

In order to determine the SDs, we numerically integrate~\eqref{eq:MSequation} by means of a custom ODE solver to extract the exact scalar power spectrum, which is then fed to \texttt{CLASS} via the {\tt external\_Pk} module.\footnote{In all cases, the numerical spectrum is subject to the constraint $\mathcal{P}_{\mathcal{R}}(k_{\star})=A_{S\star}=2.1\times 10^{-9}$~\cite{Planck:2018jri}.} Nevertheless, to scan a wide parameter space in $n_s$ and $r$ for each inflationary model, we make use of the slow-roll approximation which yields the spectrum at the {\em Planck} pivot scale $k_\star=0.05\,{\rm Mpc}^{-1}$. We consider the scalar power spectrum parametrized as usual,
\begin{equation}
		\mathcal{P}_{\mathcal{R}}(k) \;\simeq\; A_{S_\star}\left(\frac{k}{k_\star}\right)^{n_s-1 + 
		   \frac{1}{2}\frac{\mathrm dn_s}{\mathrm d\ln k}\,\ln\left(\frac{k}{k_\star}\right)+\cdots}\;,
\end{equation}
where
\begin{subequations}
\begin{align} 
\label{eq:AS}
A_{S_\star} \;&\simeq\; \frac{V(\phi_\star)}{24\pi^2\epsilon_\star M_\mathrm{P}^4}\,,\\ 
\label{eq:ns}
n_s \;&\simeq\; 1-6\epsilon_\star + 2\eta_\star\,,\\
\frac{\mathrm d\, n_s}{\mathrm d \ln k} \;&\simeq\; 16\epsilon_\star \eta_\star - 24 \epsilon_\star^2 - 2\xi_\star^2\,.
\end{align}
\end{subequations}

In the case of tensor modes with polarization states $\gamma=+,\times$, the solution of the propagation equation
\begin{equation}
\ddot{h}_{k,\gamma} + 3H \dot{h}_{k,\gamma} - \frac{k^2}{a^2}h_{k,\gamma} ~=~ 0\,,
\end{equation}
in the slow-roll approximation yields the well-known result for the tensor-to-scalar ratio
\begin{equation}\label{eq:r}
r \;:=\; \frac{A_{T_\star}}{A_{S_\star}} \;\simeq\; 16\epsilon_\star\,,
\end{equation}
where 
\begin{equation}
\sum_{\gamma=+,\times} \langle h_k h^*_{k'}\rangle ~=~ \frac{2\pi^2}{k^3}\mathcal{P}_{\mathcal{T}}(k)\,\delta(k-k')\,,
\end{equation}
and  $A_{T_\star}=\mathcal{P}_{\mathcal{T}}(k_\star)$.

\begin{figure}[t!]
	\hspace*{-12pt}
	\centering
	\includegraphics[scale=0.48]{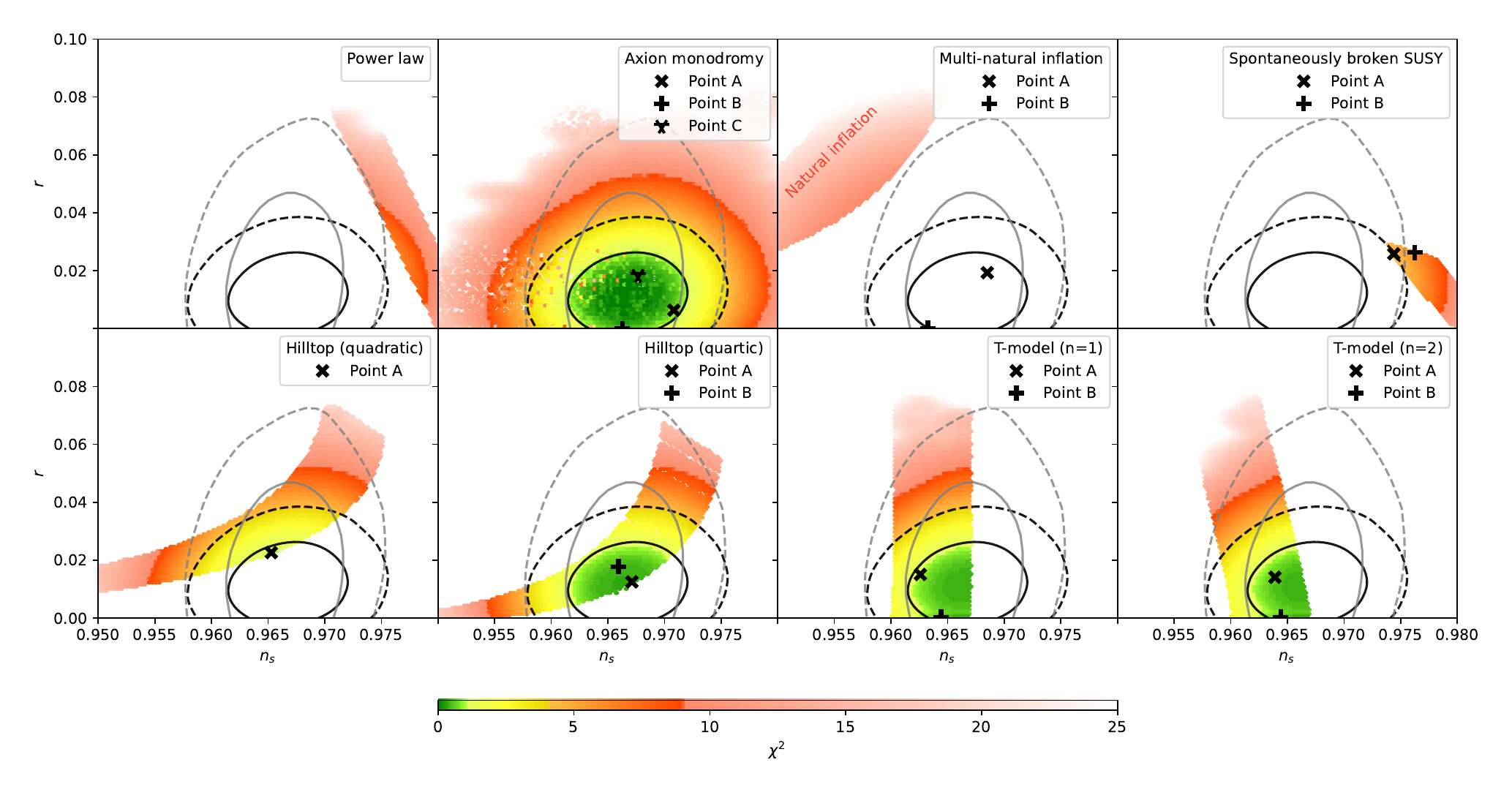}
	\caption{
	Overview of the resulting $n_s$ and $r$ for various models of single field inflation. 
	For each model, the colored area represents the accessible region in the $n_s$--$r$-plane, 
	for $50 < N_\star < 60$ $e$-folds. As usual, the green, yellow and orange regions can be
	understood as the $1\sigma$, $2\sigma$, and $3\sigma$ CL regions, while the opaque red 
	fades out until the $5\sigma$ barrier is reached. Note that $\chi^2$ includes contributions from 
	$n_s$, $r$ and the running $\frac{\mathrm{d} n_s}{\mathrm{d ln} k}$.
	The gray contour lines represent the $68\%$ and $95\%$ confidence levels from 
	{\em Planck} in combination with BK15+BAO~\cite{Planck:2018jri}, while the black 
	lines include data from BK18+BAO~\cite{BICEP:2021xfz}. Finally, the marked points
	correspond to the sample promising scenarios of the models, as discussed in
	Section~\ref{sec:SingleFieldInflation}.
	\label{fig:anisotropies_plot}}
\end{figure}

The current joint CMB constraints on the scalar tilt and tensor-to-scalar ratio are shown in Fig.~\ref{fig:anisotropies_plot}, in gray for the combined {\em Planck}+BK15+BAO data analysis~\cite{Planck:2018jri}, and in black for the {\em Planck}+BK18+BAO dataset~\cite{BICEP:2021xfz}. For the latter, the marginalized bounds correspond to $r\lesssim 0.026$ ($r\lesssim 0.039$) at the 68\% CL (95\% CL) for $n_s=0.968$, and $0.962 <n_s < 0.971$ ($0.958 <n_s < 0.975$) for $r=0.004$. Not shown in this figure is the {\em Planck} constraint on the running of the scalar index, which at the $1\sigma$ level is $\frac{\mathrm{d} n_s}{\mathrm{d}\,\mathrm{ln}\,k}=-0.0045\pm0.0067$~\cite{Planck:2018jri}.

In Fig.~\ref{fig:anisotropies_plot}, we also present a quick survey of the predictions for the inflationary observables given by the single-field models of inflation discussed in this section. As we shall shortly discuss for each case, we perform a $\chi^2$ analysis to fit the parameters of our models, evaluated at the {\em Planck} pivot scale, to the experimental CMB results. The value of $\chi^2$ includes contributions from $n_s$, $r$, and the running $\frac{\mathrm{d} n_s}{\mathrm{d}\,\mathrm{ln}\,k}$ that are summed up quadratically. 
For the contribution from the running $\alpha_s := \frac{\mathrm{d} n_s}{\mathrm{d}\,\mathrm{ln}\,k}$ we use the standard expression
\begin{equation}
\chi_{\alpha_s} ~=~ \dfrac{x_{\alpha_{s},\mathrm{model}}-x_{\alpha_{s},\mathrm{exp}}}{\sigma_{\alpha_s}}\;,
\end{equation}
where $x_{\alpha_{s},\mathrm{model}}$ and $x_{\alpha_{s},\mathrm{exp}}$ correspond to the predicted value of the model and the experimentally measured value, respectively, while $\sigma_{\alpha_s}$ denotes the standard deviation of the measured value. 
The contributions from $n_s$ and $r$, on the other hand, are determined from a two-dimensional $\chi^2$ profile constructed from the density data, which was obtained in \cite{BICEP:2021xfz} using \texttt{CosmoMC} \cite{Lewis:2002ah}.\footnote{The density data projected onto the $n_s$\;\!--\;\!$r$\;\!-plane is available within the so-called ``rns\_code'' data product on the BICEP/Keck website (\url{http://bicepkeck.org/}).}
This allows us to better factor in non-Gaussianities and correlations among the errors of $n_s$ and $r$, which would not be captured by a conventional $\chi^2$. Correlations with the errors of the running $\frac{\mathrm{d} n_s}{\mathrm{d}\,\mathrm{ln}\,k}$ are, however, not accounted for in this simple analysis.
As usual, the most acceptable regions, with up-to $3\sigma$ CL are within the green-through-orange region in each plot of Fig.~\ref{fig:anisotropies_plot}. Some promising benchmark points are indicated with special symbols; their properties shall be discussed below.

\subsection{Axion monodromy}
\label{sec:axion-monodromy}

\subsubsection{Power-law inflation}

One of the simplest proposals for slow-roll inflation is described by the power-law 
potential~\cite{Lucchin:1984yf}
\begin{equation}
\label{eq:PowerLaw}
V(\phi) ~=~ \lambda^{4-p} \phi^p \,,
\end{equation}
where the parameter $\lambda$ is fixed by the value of $A_{S_\star}$, and $p$
is supposed to fit all other observational data. The predictions for the CMB 
observables in this model are well known, and in the slow-roll approximation evaluate to
\begin{subequations}
\begin{align}
n_s ~&=~ 1-\frac{2(p+2)}{4N_{\star}+p}\,,\\
r ~&=~ \frac{16p}{4N_{\star}+p}\,.
\end{align}
\end{subequations}

As we observe in the upper left panel of Fig.~\ref{fig:anisotropies_plot}, between $N_{\star}=50$ and 60, even 
the best predictions of power-law inflation are in strong tension with current constraints 
on the spectral tilt of scalar perturbations and the tensor-to-scalar ratio~\cite{Planck:2018jri,BICEP:2021xfz}.
This can be improved if the potential that governs the inflaton dynamics includes
additional elements, as happens in axion monodromy.

\subsubsection{Improving power law by axion-like dynamics}

Axion monodromy inflation arises naturally in some string scenarios~\cite{Silverstein:2008sg,McAllister:2008hb} 
endowed with an axion-like field $\phi$ that plays the role of the inflaton.
The potential encoding some (generally field-dependent) drifting is 
given by~\cite{Flauger:2014ana}
\begin{equation}
\label{eq:AxionMonodromyPotential}
V(\phi) ~=~ \lambda^{4-p} \left[\phi^p + b\,p\,f_0\,\phi_\star^{p-1} 
            \cos\left(\dfrac{\phi^{1+p_f}}{f_0\,\phi_\star^{p_f}} + \gamma_0\right)\right]\,,
\end{equation}
which depends on the monomial power $p<1$, the axion decay constant $f_0$, the
modulation parameter $b$ (that is frequently considered to be small), the 
drifting parameter $p_f$, fixed here to $-0.7$ to maximize the spectral
distortions~\cite{Henriquez-Ortiz:2022ulz}, and the arbitrary phase $\gamma_0$.
Hence, we have four free parameters in this inflationary model to contrast 
with data. The global amplitude $\lambda$ can always be fit by comparing 
with $A_{S_\star}$ and is thus ignored here. 
The CMB observables $n_s$ and $r$ can be determined from Eqs.~\eqref{eq:ns} 
and~\eqref{eq:r}, where in this case
\begin{subequations}\label{eq:slowroll2}
\begin{eqnarray}
  \epsilon_{\star} &\simeq & \frac{p^{2}}{2} \left[ \frac{ 1-b \left(1+p_{f}\right)\sin \left(\frac{\phi_{\star} }{f_{0}}+\gamma_{0}\right)}{\phi_{\star}  + bpf_{0} \cos \left(\frac{\phi_{\star} }{f_{0}}+\gamma_{0}\right)} \right]^{2}  \,, \\
    \eta_{\star} &\simeq &  p \left[ \frac{ \left( 1+p \right)-b \left( 1+p_{f} \right)  \left[ p_{f} \sin \left(\frac{\phi_{\star} }{f_{0}}+\gamma_{0}\right) + \frac{\left( 1+p_{f} \right) \phi_{\star}}{f_0} \cos \left(\frac{\phi_{\star} }{f_{0}}+\gamma_{0}\right)  \right] }{\phi^{2}_{\star}+bpf_{0} \phi_{\star}  \cos \left(\frac{\phi_{\star} }{f_{0}}+\gamma_{0}\right)} \right] \,.
\end{eqnarray}
\end{subequations}
upon evaluation of Eq.~\eqref{eq:slowroll0} with the potential~\eqref{eq:AxionMonodromyPotential}.

The possible ranges of values for the CMB observables $(n_s,r)$ for the axion monodromy model are shown in the correspondingly labeled panel of Fig.~\ref{fig:anisotropies_plot}. It is worth noting that, upon continuous variation of the model parameters, it is possible to cover almost entirely the $1\sigma$, $2\sigma$ and $3\sigma$ CL {\em Planck}+BK18+BAO regions.
However, the modulation of the inflaton potential (\ref{eq:AxionMonodromyPotential}) by the periodic term induces an oscillatory behavior on top of the near scale-invariant curvature power spectrum, leading to a potentially large local running of $n_s$.

In order to simplify the analysis, we focus on three phenomenologically viable choices of the model parameters, which are shown in Table~\ref{tab:tabAM1}, labeled as points A,B,C. For simplicity, the parameter $p_f$ and the number of $e$-folds at the {\em Planck} pivot scale $N_{\star}$ are fixed to $-0.7$ and 57.5, respectively, for the three cases. Point A, with the largest decay constant $f_0$ and shown as the `$\times$' point in Fig.~\ref{fig:anisotropies_plot}, leads to the largest value for the scalar tilt, $n_s=0.971$, and a small tensor-to-scalar ratio, $r=0.006$, comparable to scenarios with asymptotically flat potentials, such as the T-model (see Section~\ref{sec:Tmodel}). Point B, with intermediate $f_0$ and the largest value for $b$, nearly matches the central value of $n_s$. Notably, in this case we obtain the lowest tensor-to-scalar ratio, $r\sim\mathcal{O}(10^{-10})$, far below current and near-future experimental detection thresholds. Finally, point C, with the smallest $f_0,b$, is also near the center of the preferred $n_s$ range, with $r=0.02$, within the detection capabilities of next generation CMB observatories~\cite{SimonsObservatory:2018koc,LiteBIRD:2022cnt,Abazajian:2019eic}.

\begin{table}[t!]
	\centering
	\begin{tabular}{l|llllll|lll|l}
		\toprule
		Point & $p$ & $b$ & $f_0$  & $\gamma_0$ & $p_f$ & $N_{\star}$ & $n_s$ & $r$ & $\frac{\mathrm{d}\,n_s}{\mathrm{d\,ln}\,k}$ & $\chi^2$ \\ \midrule
		Point A & $0.0919$ & $0.0231$  & $0.01$     & $1.68$  & $-0.700$ & $57.5$ & $0.971$ & $0.0064$ & $-0.004$ & $0.7$ \\
		Point B & $0.02$   & $3.33$    & $0.0036$   & $1.87$  & $-0.700$ & $57.5$ & $0.966$ & $5\cdot 10^{-10}$ & $-0.004$ & $0.5$ \\
		Point C & $0.266$  & $0.00134$ & $0.000918$ & $0.476$ & $-0.700$ & $57.5$ & $0.968$ & $0.0185$ & $-0.006$ & $0.2$ \\\bottomrule
	\end{tabular}
	\caption{Phenomenologically viable sample points for axion monodromy, and their associated predictions for the CMB observables. See Fig.~\ref{fig:anisotropies_plot}.\label{tab:tabAM1}}
\end{table}

\begin{figure}[t!]
	\includegraphics[scale=0.26]{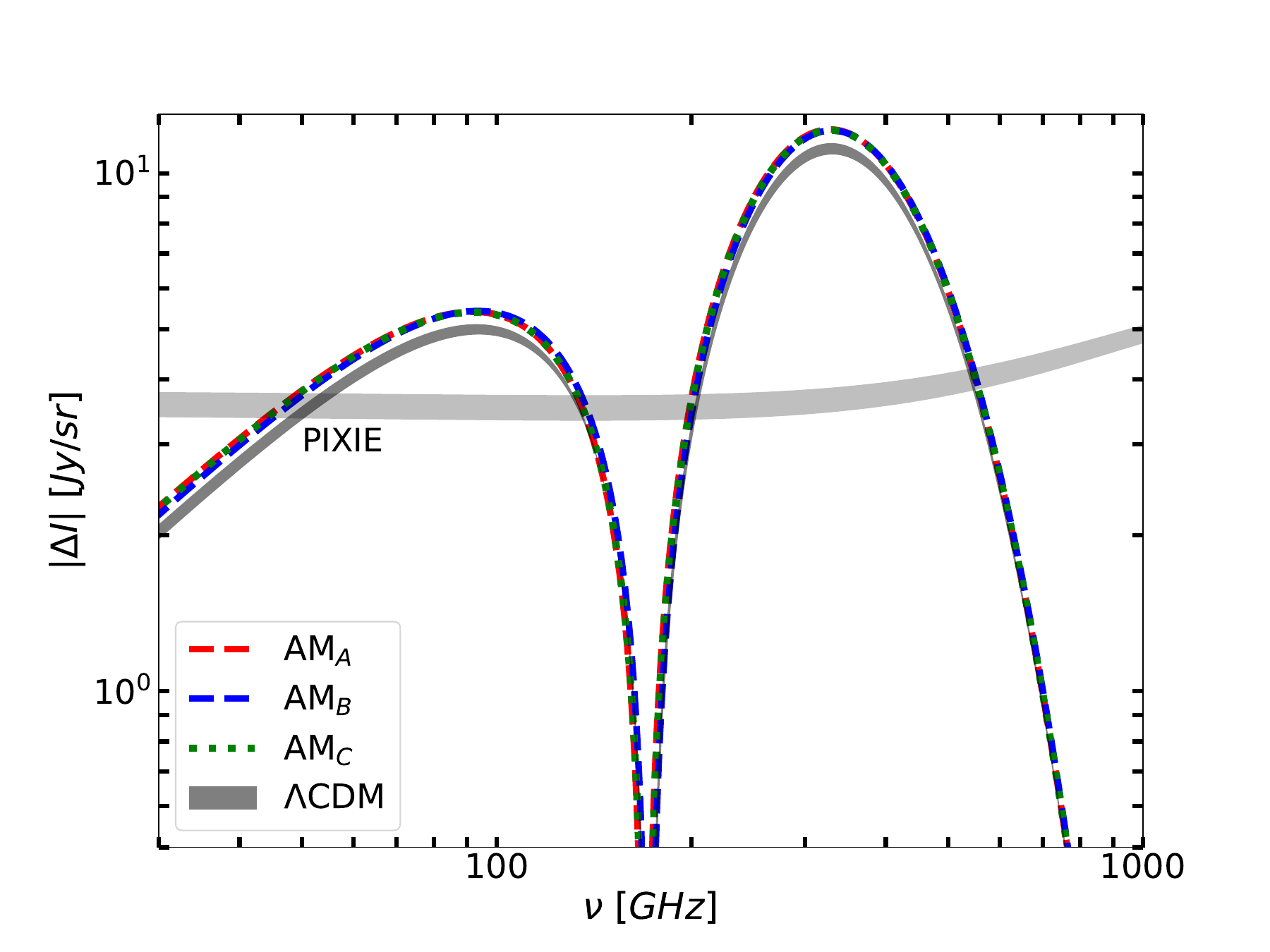}
	\includegraphics[scale=0.26]{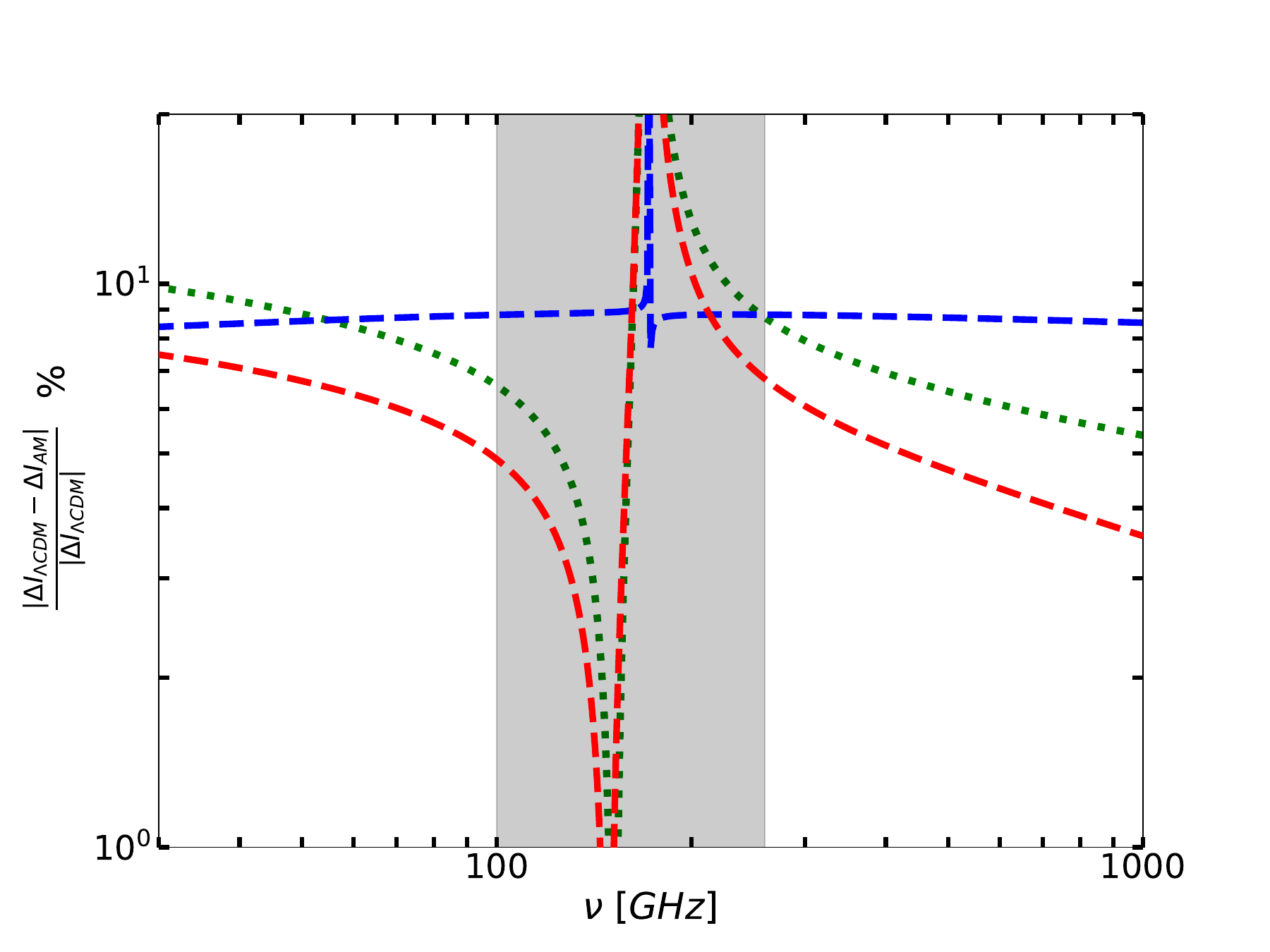}
	\caption{Left:~predictions for the distortion $\Delta I$ of the photon intensity~\eqref{eq:sdn2} in axion monodromy inflationary models at the benchmark points defined in Table~\ref{tab:tabAM1}, contrasted against the $\Lambda$CDM prediction. Three dashed curves are shown, corresponding to AM$_{A}$ (axion-monodromy point A), AM$_{B}$ (point B), AM$_{C}$ (point C), together with the $\Lambda$CDM prediction, depicted as the dark gray region. We further show the sensitivity of future PIXIE (curve adapted from~\cite[fig.~9]{Chluba:2019nxa}). Right: percentage difference between the axion-monodromy prediction $\Delta I_\mathrm{AM}$ in the cases of interest A,B,C and $\Delta I_{\Lambda\mathrm{CDM}}$. For frequencies outside the range $100\,\mathrm{GHz}\lesssim\nu\lesssim250\,\mathrm{GHz}$, the difference w.r.t.\ the fiducial signal is about 10\%.\label{fig:AMresults}}
\end{figure}

To determine the predictions of axion monodromy for SDs signals, we made use of {\tt CLASS} to numerically compute the total distortion $\Delta I$ of the photon intensity as a function of the frequency $\nu$, for the benchmark points shown in Table~\ref{tab:tabAM1}. The left panel of Fig.~\ref{fig:AMresults} contrasts the corresponding predictions, labeled also A,B,C and shown as dashed curves, against the $\Lambda$CDM prediction, shown as a dark gray region. This region is obtained by varying the spectral tilt within its observed 1-$\sigma$ region, considering {\em Planck}'s result $n_s=0.9665\pm0.0038$~\cite[Eq.~(14)]{Planck:2018jri}.
Here the distortion is shown in units of $\mathrm{Jy}/\mathrm{sr}=10^{-26}\, \mathrm{W}\,\mathrm{m}^{-2}\mathrm{Hz}^{-1}\mathrm{sr}^{-1}$. Also in the left panel, the projected sensitivity of the PIXIE experiment is displayed as the light gray shaded region. In all three cases, the distortion has a larger amplitude than the $\Lambda$CDM case, by up to $\sim$10\% in the physically relevant region, $\nu\lesssim100$\,GHz and $\nu\gtrsim250$\,GHz.\footnote{With this conservative choice, we avoid the region $100\,\mathrm{GHz}\lesssim\nu\lesssim250\,\mathrm{GHz}$, where the difference between the signals of our models and the fiducial $\Lambda$CDM is strongly reduced, and the SDs signal itself aproaches zero, rendering it unavailable to any of the conceived probes and hence physically less relevant.} This is more clearly visible in the right panel of Fig.~\ref{fig:AMresults}, where the difference  $|\Delta I_{\Lambda\mathrm{CDM}}-\Delta I_\mathrm{AM}|$ is shown as a percentage of the $\Lambda$CDM result. Note that a meaningful comparison of the SDs produced at each axion-monodromy point can be only achieved if the value of $n_s$ taken for the $\Lambda$CDM spectrum coincides with the one produced by the model, as given in Table~\ref{tab:tabAM1}.

\subsection{Hilltop models}

After the original proposal for inflation ({\em old inflation}) was abandoned due to its impossibility of reconciling a sufficiently long period of accelerated expansion with a graceful exit~\cite{Guth:1980zm,Guth:1982pn}, inflationary models were soon after introduced, which relied on a second-order transition from a symmetric to a broken phase~\cite{Albrecht:1982wi,Linde:1981mu}. In such ``new'' inflationary models, symmetry restoration at high temperatures would naturally drive the inflaton to zero due to the appearance of a minimum with $V(0):= \Lambda^4 > 0$, as $m_{\phi} \propto T$. Stuck in the false vacuum, inflation would begin, cooling down the universe, and turning the origin into a local maximum, from which the inflaton would slowly roll to the true, low-energy minimum. Originally, these {\em hilltop} models were based on the Coleman-Weinberg correction to the vacuum potential, which is incompatible with the  constraints by {\em Planck}~\cite{Barenboim:2013wra}. Nevertheless, generic hilltop models with a potential of the form
\begin{equation}
\label{eq:Vhilltop}
V(\phi) ~=~ \Lambda^4 \left(1-\dfrac{\phi^p}{\mu^p} + \dots\right)\;,
\end{equation}
can still be in agreement with current CMB constraints. In this potential the ellipsis denotes the extra terms in $V(\phi)$ necessary to complete it from below, in order to have a minimum with a phenomenologically allowed cosmological constant. For example, the quadratic ($p=2$) hilltop model can be completed by a simple symmetry breaking potential $V(\phi)\propto (1-\phi^2/\mu^2)^2$~\cite{Olive:1989nu}.

We scan the allowed parameter space for the quadratic and the quartic ($p=4$) hilltop models, without any assumptions regarding the completion of the potential~\eqref{eq:Vhilltop}. In the slow-roll approximation, the inflaton field at the horizon exit of the pivot scale $k_{\star}$ can be related with the number of $e$-folds as
\begin{equation}\label{eq:Nstarphiend}
N_{\star} \;=\; N(\phi_{\rm end}) - N(\phi_{\star})\,,
\end{equation}
where
\begin{equation}
N(\phi) \;=\; -\int^{\phi} \frac{d\phi'}{\sqrt{2\epsilon(\phi')}} \;=\; \begin{cases}
\dfrac{\phi ^2}{4 } - \dfrac{\mu^2}{2} \ln \left(\dfrac{\phi }{\mu }\right)\,, & p=2\,,\\[10pt]
\dfrac{\phi^2}{8} + \dfrac{1}{8}\left(\dfrac{\mu^2}{\phi }\right)^2\,, & p=4\,,
\end{cases}
\end{equation}
and the inflaton value at the end of inflation can be obtained from the solution of $\epsilon(\phi_{\rm end})= 1$. In all cases, the slow-roll parameters are given by
\begin{subequations}
\begin{align}
\epsilon(\phi) \;&\simeq\;  \frac{p^2(\phi/\mu)^{2p-2}}{2\mu^2(1-(\phi/\mu)^p)^2}\,,\\
\eta(\phi) \;&\simeq\; \frac{p(p-1)(\phi/\mu)^{p-2}}{\mu^2\left(1-(\phi/\mu)^p\right)}\,. 
\end{align}
\end{subequations}
Together with the (numerical) solution for $\phi_{\star}$, the CMB observables may be evaluated from Eqs.~\eqref{eq:ns} and~\eqref{eq:r}. 

For the quadratic hilltop model, we scan over the range $9\lesssim \mu\lesssim 150$ and $50<N_{\star}<60$. The results are  
shown in the bottom left panel of Fig.~\ref{fig:anisotropies_plot}. The lowest $\chi^2$ value corresponds to considering $\mu\simeq 13$ and $N_{\star}=60$, and is labeled as Point A. In this case, the scalar tilt and the tensor-to-scalar ratio are given in Table~\ref{table:hilltop}, and are barely inside the 68\% CL level region of the {\em Planck}+BK18 constraints.
Larger values of $N_\star$ may lead to a better compatibility, but they would require some non-trivial reheating dynamics. The running of the scalar tilt is also in the experimentally preferred region for this parameter space point, as shown in Table~\ref{table:hilltop}. 

\begin{table}[t!]
	\centering
\begin{tabular}{l|ll|lll|l}
	\toprule
	Point & $\mu$ & $N_{\star}$ & $n_s$ & $r$ & $\frac{\mathrm{d}\,n_s}{\mathrm{d\,ln}\,k}$ & $\chi^2$ \\ \midrule
	quadratic A & $13.0$ & $60.0$ & $0.965$ & $0.0227$ & $-0.000345$ & $1.05$ \\
	quartic A & $16.0$ & $60.0$ & $0.967$ & $0.0125$ & $-0.000456$ & $0.38$ \\
	quartic B & $18.0$ & $55.0$ & $0.966$ & $0.0178$ & $-0.000524$ & $0.50$ \\\bottomrule
\end{tabular}
\caption{Sample points for a hilltop scenario with quadratic and quartic potentials, and their corresponding CMB observables. See also Fig.~\ref{fig:anisotropies_plot}. \label{table:hilltop}}
\end{table}

For the $p=4$ case, we scan over the range $1\lesssim \mu\lesssim 150$ and $50<N_{\star}<60$. The resulting values of $n_s$ and $r$ are shown in the second panel in the bottom of Fig.~\ref{fig:anisotropies_plot}. The preferred range of parameters, shown in green and sitting comfortably inside the $1\sigma$ {\em Planck}+BK18 contour, is centered around $\mu\simeq 16$ for $N_{\star}\gtrsim 53$. In this region we have selected two sample points, denoted as `$\times$' (point A) and `$+$' (point B), whose CMB signatures can be explicitly found in Table~\ref{table:hilltop}. For these points, the constraint on the running of $n_s$ is also satisfied. 

\begin{figure}[t!]
	\includegraphics[scale=0.26]{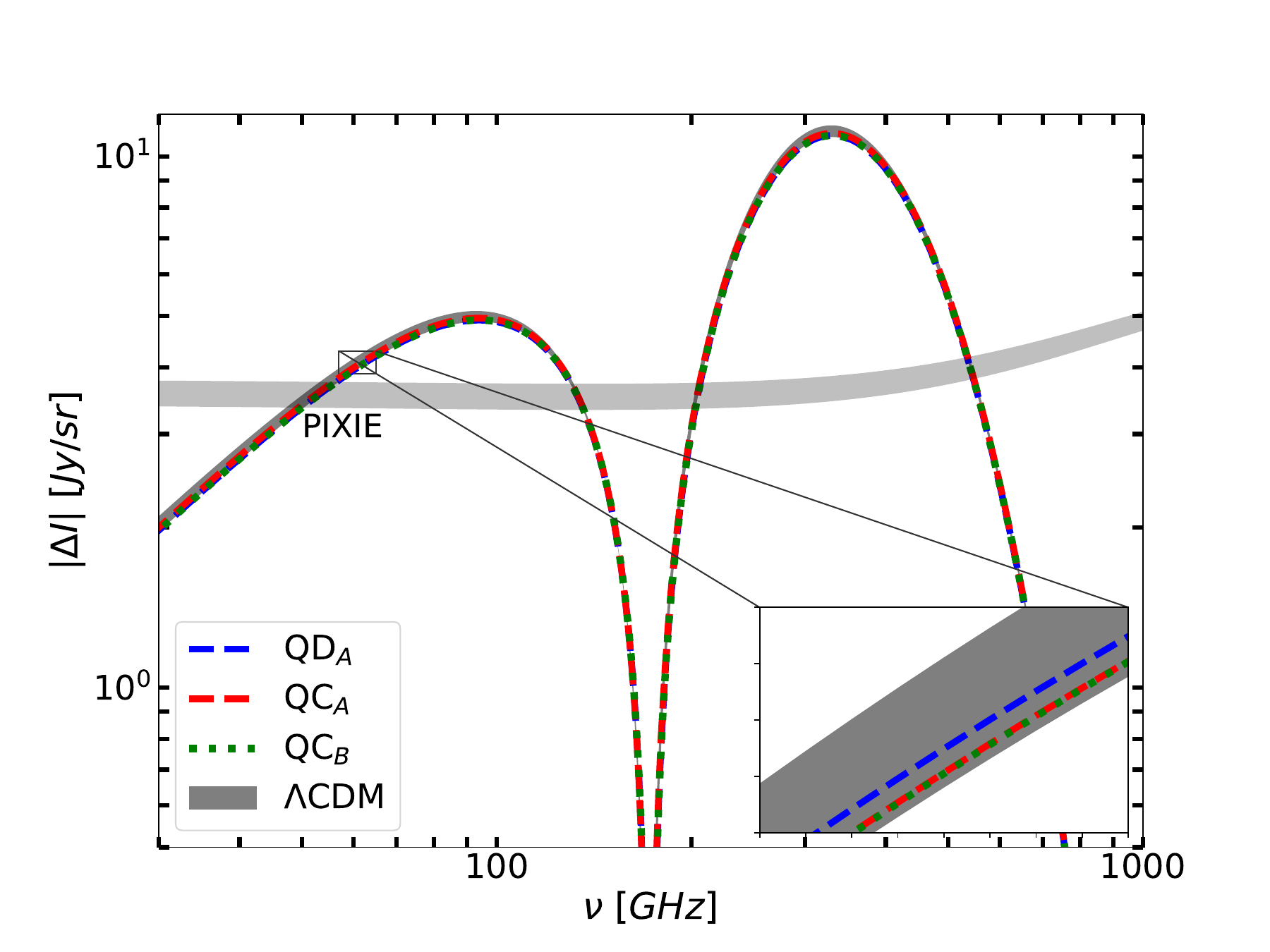}
	\includegraphics[scale=0.26]{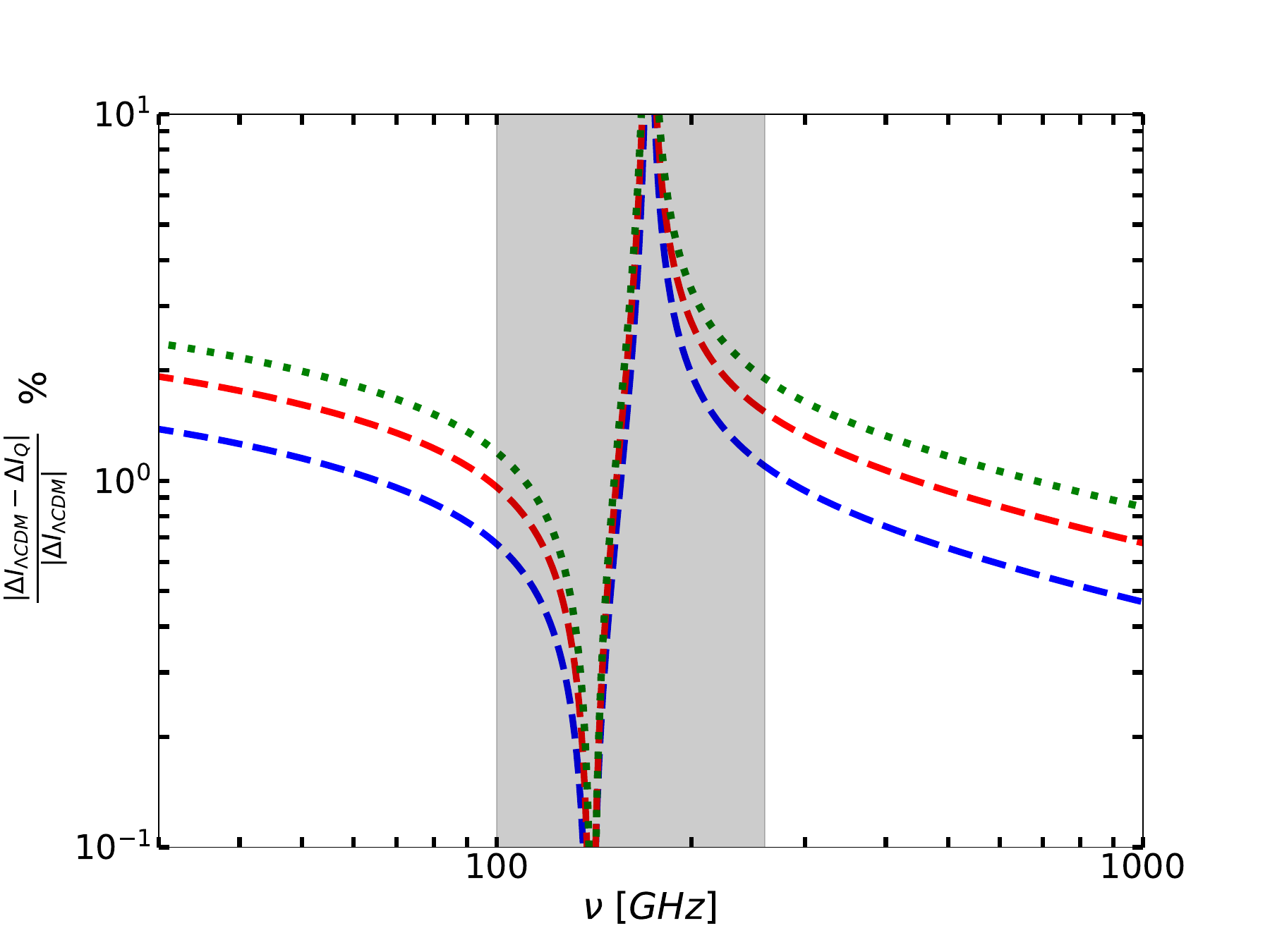}
	\caption{Left: predicted $\Delta I$ in hilltop models for the benchmark points defined in Table~\ref{table:hilltop}, contrasted against the $\Lambda$CDM prediction. The dashed curves result from the parameters of the quadratic A (QD$_{A}$), quartic A (QC$_{A}$) and quartic B (QC$_{B}$) points, while the $\Lambda$CDM prediction is shown as a dark gray region. As in Fig.~\ref{fig:AMresults}, we also display the sensitivity of future PIXIE. Right: percentage difference between the hilltop prediction $\Delta I_\mathrm{Q}$ and $\Delta I_{\Lambda\mathrm{CDM}}$. For frequencies outside the range $100\,\mathrm{GHz}\lesssim\nu\lesssim250\,\mathrm{GHz}$ the difference w.r.t.\ the fiducial signal is about 1\%-2\%.}
	\label{fig:sdhiltop}
\end{figure}

Fig.~\ref{fig:sdhiltop} shows the predictions for the amplitude $\Delta I$ of SDs associated with the sample points in Table~\ref{table:hilltop}, contrasted against the $\Lambda$CDM prediction, and compared to the sensitivity of the PIXIE experiment. The left panel displays the absolute contribution to SDs, with the results for the quadratic A (QD$_{A}$), quartic A (QC$_{A}$) and quartic B (QC$_{B}$) sample points given by the blue, red and green  curves, respectively. The right panel shows the relative difference of the distortions with respect to the $\Lambda$CDM result. We note that the SDs can differ only by up to about 1\%-2\% approximately in the physically relevant region, $\nu\lesssim100$\,GHz and $\nu\gtrsim250$\,GHz.

\subsection{Multi-natural inflation}
\label{sec:multinatural}

Upon the advent of large field, `chaotic' inflation models, transplanckian excursions of the inflaton field during the slow-roll phase have been a common feature of a large class of models~\cite{Linde:1983gd}. Maintaining the flatness of the potential at such large-field values is however challenging, as this generically requires the presence of small parameters, unstable against radiative corrections. Among the proposals to alleviate this so-called $\eta$-problem, the identification of the inflaton with an axion-like field enjoying 
an approximate shift symmetry, as in the case of axion monodromy, is found to be 
a {\em natural} solution for this problem~\cite{Freese:1990rb,Adams:1992bn}. As before, in its simplest realization the inflaton potential is sinusoidal, 
\begin{equation}
V(\phi) \;=\; \Lambda^4 (1+\cos(\phi/f))\,,
\end{equation} 
originated from non-perturbative effects which break the shift symmetry. Nevertheless, in its single field realization, the model is disfavored at the $3\sigma$ level in the phenomenologically allowed region. The third, top panel of Fig.~\ref{fig:anisotropies_plot} shows the predictions for the CMB observables for natural inflation in the parameter range $50\leq N_{\star}<60$, and $4\leq f\leq 100$, contrasted with the current {\em Planck}+BK constraints.
The lowest $\chi^2$ is $9.1$.

A way around these difficulties is to assume multiple sources for the breakdown of the shift symmetry of the inflaton~\cite{Takahashi:2013tj}. If this is the case, then the resulting potential could have more than one oscillatory contribution. In the notation of~\cite{Czerny:2014wza}, for two sinusoidal terms, these {\em multi-natural} models are parametrized in terms of the scalar potential
\begin{equation}
V(\phi) ~=~ \Lambda^4 \left( - \cos\left(\dfrac{\phi}{f}\right) - 
B\,\cos\left(\dfrac{\phi}{A\,f}+\theta\right) + C\right)\;,
\end{equation}
where $C$ is a constant that shifts the potential. Here we choose its value to eliminate a potentially large cosmological constant, i.e.\ $V(\phi_\mathrm{min}) = 0$ at the minimum. It is in addition worth noting that small deviations of $C$ have a significant impact on the slow-roll parameters and therefore change the resulting spectral tilt $n_s$ and tensor-to-scalar ratio $r$. Such a shift in the potential cannot be absorbed by a redefinition of the remaining parameters. We also note that, in some special limits, the potential of this model reduces to the hilltop scenario.

The need to adjust the value of $C$ for each given value of the rest of the parameters of the model makes a full parameter-space scan for this model computationally intensive. Additionally, the relatively complicated form for the slow-roll parameters $\epsilon$ and $\eta$, and the lack of an analytical closed form expression for $N_{\star}(\phi)$ prevent us from writing explicit expressions for the CMB observables. Hence, we focus on two particular sample points, listed in Table~\ref{tab:MNI} and shown in the third top panel of Fig.~\ref{fig:anisotropies_plot}. Both choices fall within the $1\sigma$ {\em Planck}+BK18 CL region, with point A leading to a smaller deviation from scale invariance, and a much larger amplitude for tensors, compared with point B. 

\begin{table}[t!]
	\centering
	\begin{tabular}{l|llllll|lll|l}
		\toprule
		Point & $A$ & $B$ & $C$ & $f$ & $\theta$ & $N_{\star}$ & $n_s$ & $r$ & $\frac{\mathrm{d}\,n_s}{\mathrm{d\,ln}\,k}$ & $\chi^2$ \\ \midrule
		Point A & $0.055$ & $0.00178$ & $1.00156$ & $4.35$ & $0.16\pi$ & $57.5$ & $0.968$ & $0.019$ & $-0.004$ & $0.3$ \\
		Point B & $0.22$ & $0.14$ & $1.138925$ & $2.7$ & $0.1$ & $57.5$ & $0.963$ & $0.00015$ & $-0.002$ & $1.0$ \\\bottomrule
	\end{tabular}
	\caption{Sample points for a multi-natural inflation scenario and their CMB predictions. See also Fig.~\ref{fig:anisotropies_plot}.}
	\label{tab:MNI}
\end{table}

\begin{figure}[t!]
	\includegraphics[scale=0.26]{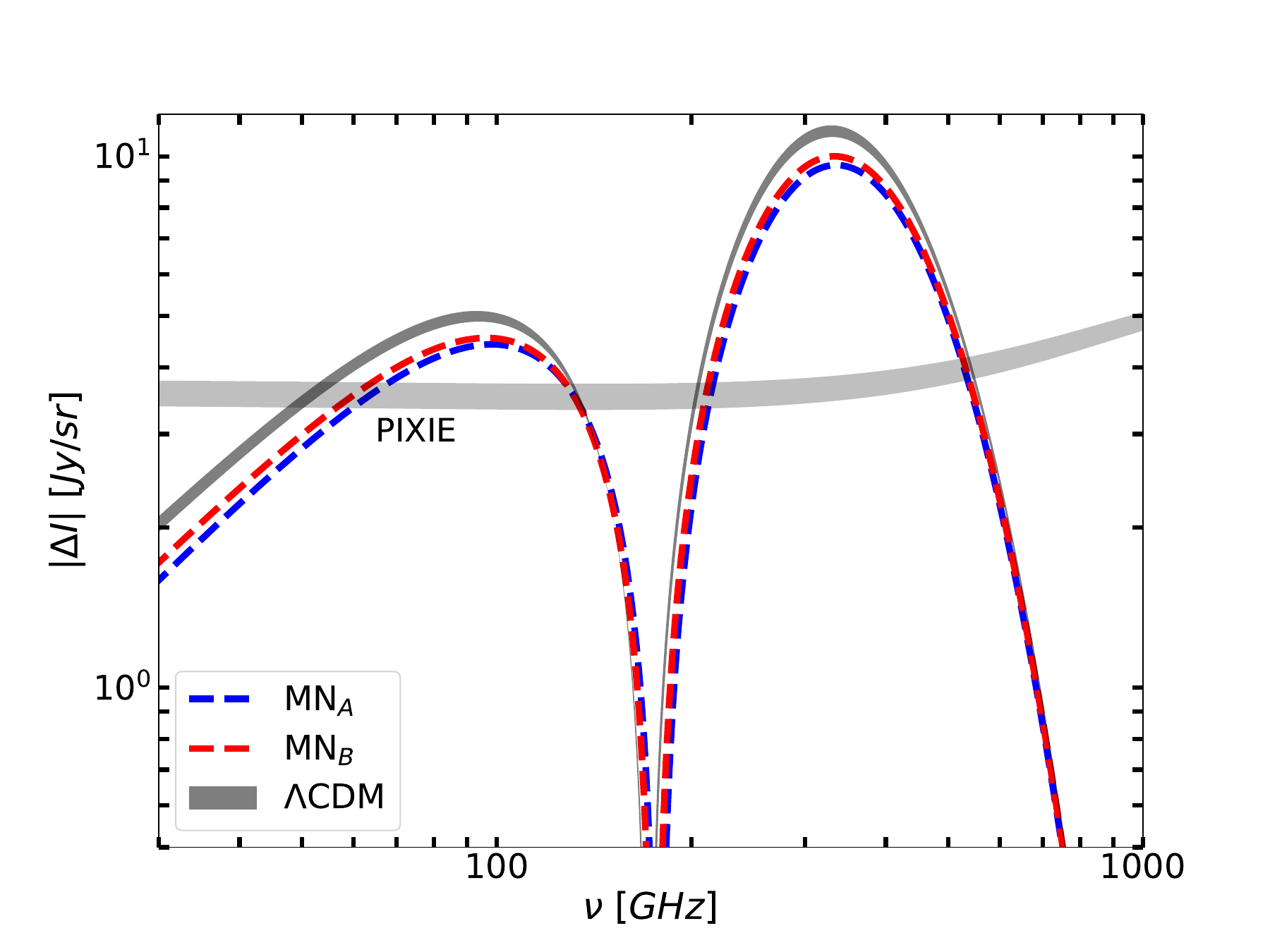}
	\includegraphics[scale=0.26]{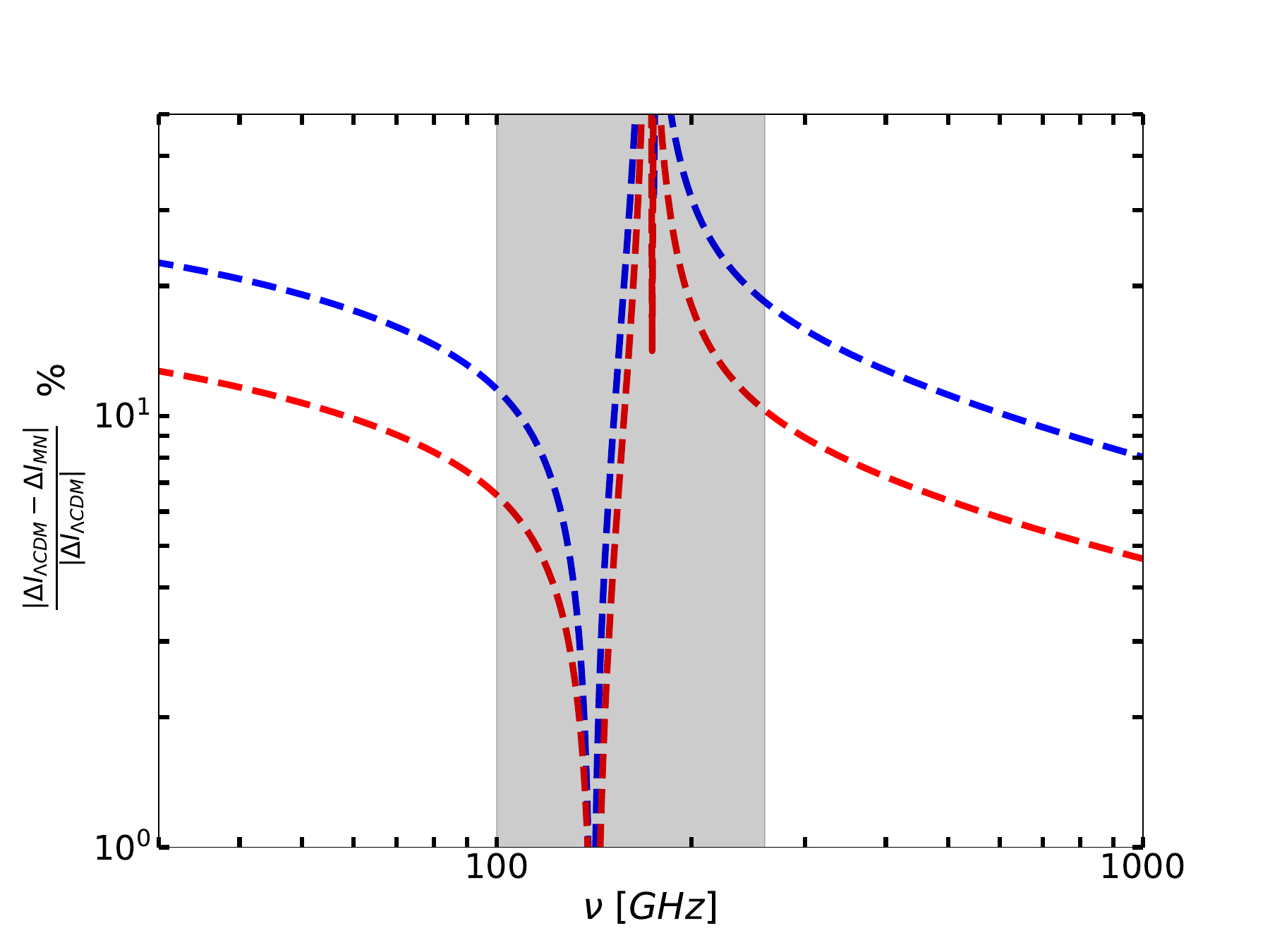}
	\caption{Left: contribution to $\Delta I$ for the two realizations of multi-natural inflation of Table~\ref{tab:MNI}, points A (MN$_{A}$)  and B (MN$_{B}$) (dashed), contrasted against the $\Lambda$CDM prediction (dark gray region) and the sensitivity of future PIXIE (gray, shaded). Right: percentage difference between the multi-natural $\Delta I_\mathrm{MN}$ signal and $\Delta I_{\Lambda\mathrm{CDM}}$. For frequencies outside the range $100\,\mathrm{GHz}\lesssim\nu\lesssim250\,\mathrm{GHz}$ the difference w.r.t.\ the fiducial signal is about 10\%-20\%.}
	\label{fig:sdmn}
\end{figure}

The left panel of Fig.~\ref{fig:sdmn} displays the amplitude of the spectral distortion for multi-natural models,  $\Delta I_\mathrm{MN}$, for the sample points in Table~\ref{tab:MNI}, shown as the red and blue dashed curves. Contrasted against the $\Lambda$CDM, the resulting signal of SDs is suppressed by about 10\%-20\% in the relevant region $\nu\lesssim100$\,GHz and $\nu\gtrsim250$\,GHz. This is more explicitly conveyed  in the right panel of Fig.~\ref{fig:sdmn}. 

\subsection{Spontaneously broken SUSY}

Another class of inflationary models arises from supersymmetric grand unified constructions that avoid the need to enlarge the Higgs sector of the Standard Model. Based on the gauge symmetry (of so-called trinification models) $\SU3_c\x\SU3_L\x\SU3_R$, one-loop corrections due to spontaneous supersymmetry (SUSY) breaking can induce a potential for the inflaton field of the form~\cite{Dvali:1994ms}
\begin{equation}
V(\phi) ~=~ \Lambda^4 \left( 1 + \beta\,\ln\phi+ \dots\right)\,.
\end{equation}
Here the ellipsis corresponds to terms that become important after slow roll ends, and which lead to the spontaneous breakdown of the gauge symmetry in a hybrid-like scenario. 

Without an explicit reheating mechanism, we are left with two free parameters, $\beta$ and the number of $e$-folds $N_{\star}$. Similarly to the hilltop scenarios, we can relate $N_{\star}$ with the inflaton field at horizon crossing by means of equation (\ref{eq:Nstarphiend}), where in this case
\begin{equation}
N(\phi) \;\simeq\; \frac{\phi^2}{4 \beta }  \left(\beta-2 -2 \beta  \ln \phi\right)\,,
\end{equation}
and by numerical determination of $\phi_{\rm end}$, as the condition $\ddot{a}=0$ leads to transcendental equations. In turn, the slow-roll parameters at horizon crossing will be given by
\begin{subequations}
\begin{align}
\epsilon_{\star}\;&\simeq\; \frac{\beta ^2 }{2 \phi ^2 \left(1 + \beta  \ln \phi \right)^2}\,,\\
\eta_{\star} \;&\simeq\; -\frac{\beta }{2 \phi ^2 \left(1 + \beta  \ln \phi \right)}\,,
\end{align}
\end{subequations}
from which $n_s$ and $r$ can be determined from Eqs.~\eqref{eq:ns} and~\eqref{eq:r}, as usual.

The results for our CMB parameter scan are shown in the upper right panel of Fig.~\ref{fig:anisotropies_plot}. We consider the nominal $50\leq N_\star\leq 60$ range, and $3\times 10^{-3}\lesssim \beta\lesssim 10^2$. For this model, in all the domain, the $\chi^2$ statistical significance is never better than $4.2$. We see that only a small corner of the parameter space, with small $N_\star$ and large $\beta$, lies within the 95\% CL region of {\em Planck} 2018 data. An even smaller piece of the parameter space lies inside the $2\sigma$ contours of the combined {\em Planck}+BK18 analysis. We have selected one sample point in this compatibility region, marked as `$\times$', and another sample point, marked as `$+$', outside the larger 2018 $2\sigma$ window. Their corresponding observables are displayed in Table~\ref{tab:SUSY}.

\begin{table}[t!]
	\centering
	\begin{tabular}{l|ll|lll|l}
		\toprule
		Point & $\beta$ & $N_{\star}$ & $n_s$ & $r$ & $\frac{\mathrm{d}\,n_s}{\mathrm{d\,ln}\,k}$ & $\chi^2$ \\ \midrule
		Point A & $2$ & $50.0$ & $0.974$ & $0.0258$ & $-0.000521$ & $4.2$ \\
		Point B & $100$ & $55.0$ & $0.976$ & $0.0263$ & $-0.000440$ & $6.2$ \\\bottomrule
	\end{tabular}
	\caption{Sample points for inflationary models based on spontaneously broken SUSY and their CMB predictions. See also Fig.~\ref{fig:anisotropies_plot}.
	\label{tab:SUSY}}
\end{table}
\begin{figure}[t!]
	\includegraphics[scale=0.26]{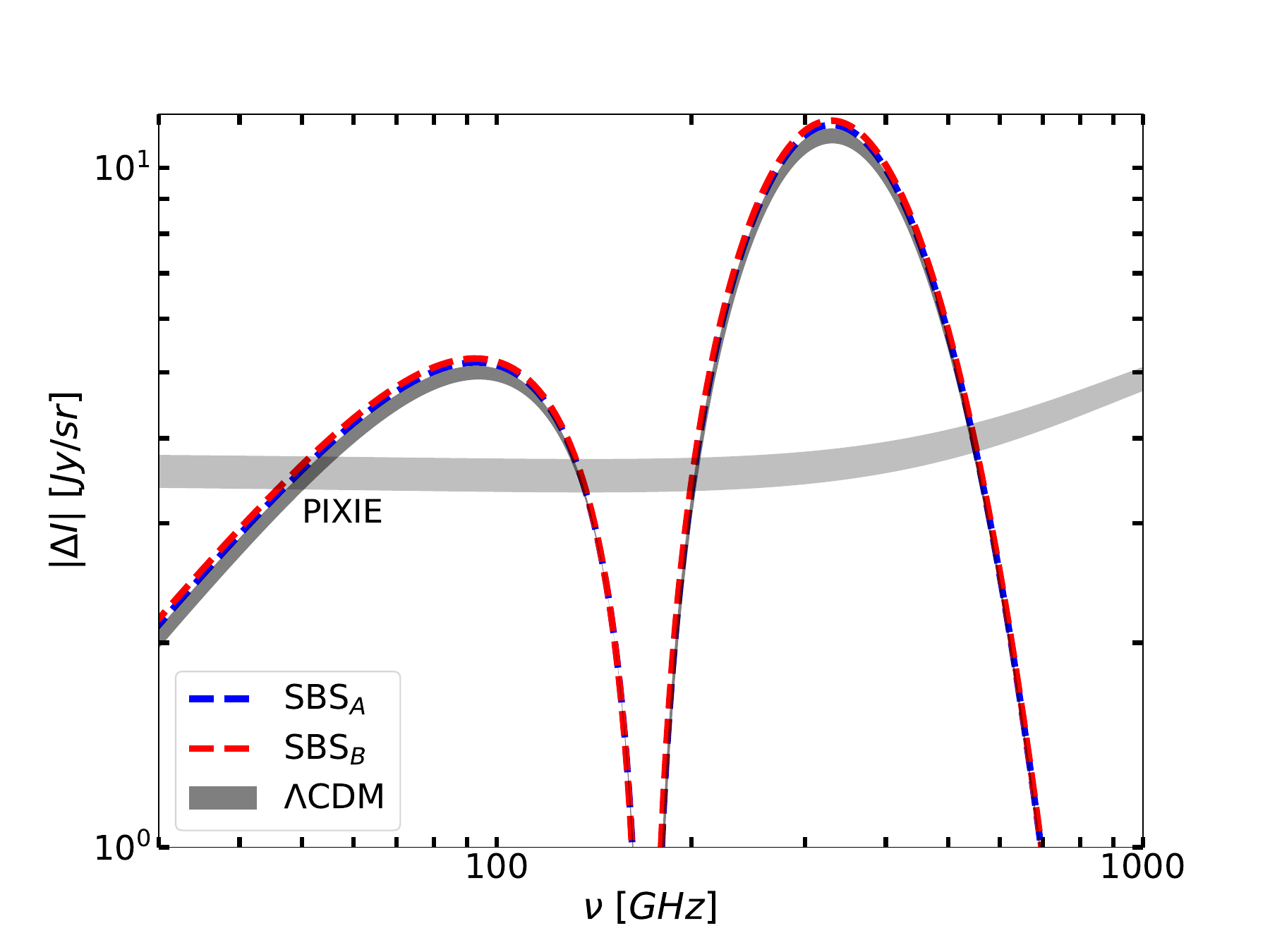}
	\includegraphics[scale=0.26]{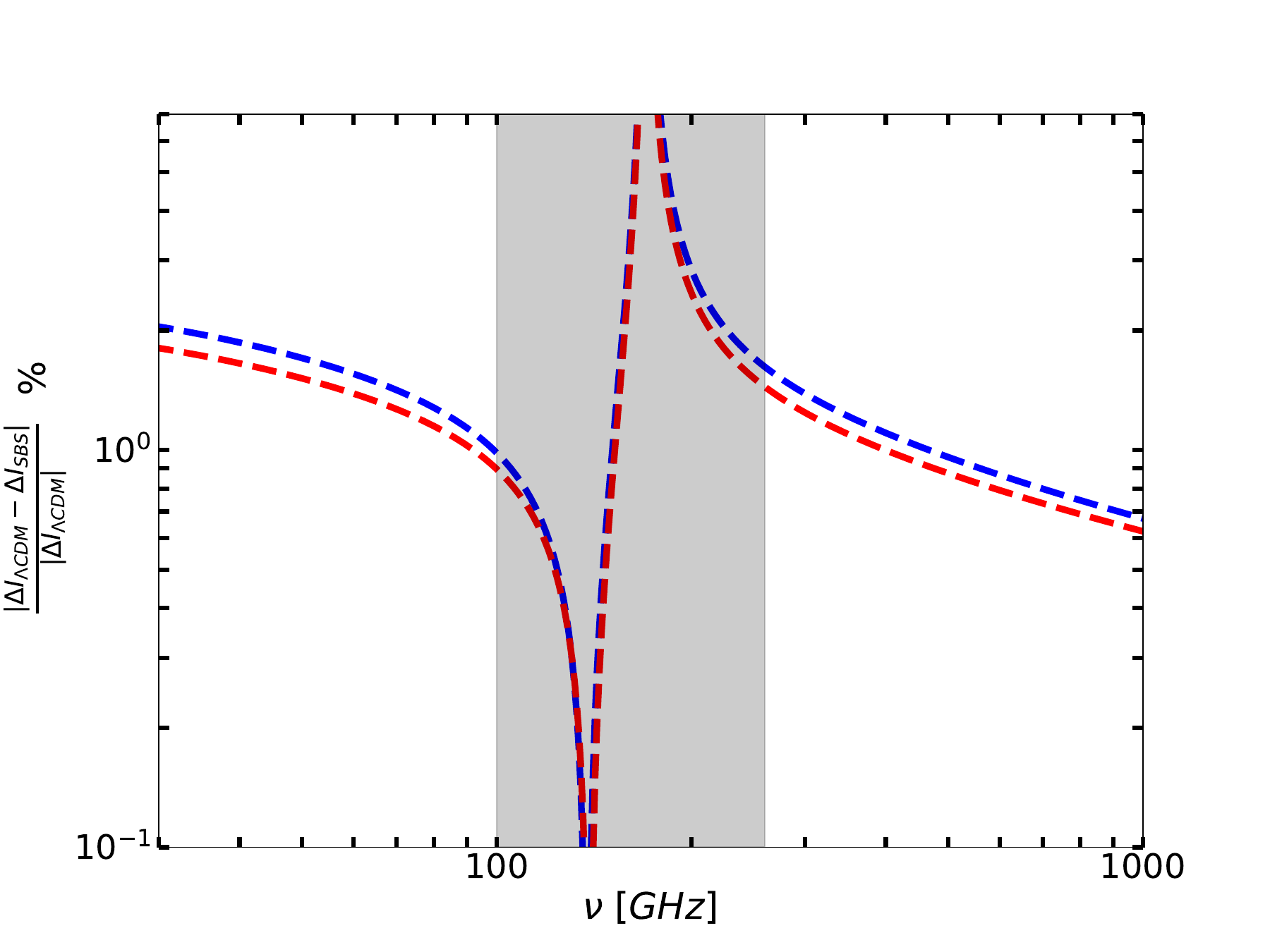}
	\caption{Left: predicted $\Delta I$ for the spontaneously broken SUSY models defined by the data in Table~\ref{tab:SUSY}, contrasted against the $\Lambda$CDM prediction. Two dashed curves are shown, corresponding to point A (SBS$_{A}$) and point B (SBS$_{B}$), together with the $\Lambda$CDM prediction (dark gray region) and the sensitivity of future PIXIE. Right: percentage difference between the spontaneously broken SUSY prediction for $\Delta I_\mathrm{SBS}$ and $\Delta I_{\Lambda\mathrm{CDM}}$. For frequencies outside the range $100\,\mathrm{GHz}\lesssim\nu\lesssim250\,\mathrm{GHz}$ the difference w.r.t.\ the fiducial signal is about 1\%-2\%.}
	\label{fig:sdsusy}
\end{figure}

The predictions for SDs from the spontaneously-broken SUSY inflationary scenarios, $\Delta I_\mathrm{SBS}$, defined by the points in Table ~\ref{tab:SUSY}, are shown in Fig.~\ref{fig:sdsusy}, contrasted against the $\Lambda$CDM prediction and the sensitivity of the PIXIE experiment. The left panel shows the prediction for point A (SBS$_{A}$) and point B (SBS$_{B}$) as the dashed curves for the absolute distortion. In the right panel we display the difference as a percentage of the $\Lambda$CDM result. We see that they can differ by up to about 2\% approximately in the physically relevant region, $\nu\lesssim100$\,GHz and $\nu\gtrsim250$\,GHz.

\subsection{T-model}\label{sec:Tmodel}

A class of inflationary models that easily accommodates the {\em Planck}+BK upper limit on $r$, in addition to being compatible with the scalar spectrum constraints, are large-field plateau models. Of particular interest are the family of inflationary $\alpha$-attractors, which include the $\alpha$-Starobinsky models (also known as E-models)~\cite{Ellis:2013nxa,Kallosh:2013yoa,Ellis:2019bmm}, and the T-models~\cite{Kallosh:2013maa}. The latter are described by a potential of the form
\begin{equation}\label{eq:VTmodel}
V(\phi) ~=~ \Lambda^4~ \mathrm{tanh}^{2n}\left(\dfrac{\phi}{\sqrt{6\,\alpha}}\right)\;,
\end{equation}
which is asymptotically flat at large field values, $\phi\gg 1$. The parameter $\alpha$ determines the flatness of the plateau at the moment of the horizon exit of the CMB pivot scale. More precisely, increasing $\alpha$ decreases the curvature of the potential at $\phi=\phi_{\star}$. Since $V(\phi)\propto \phi^{2n} $ for $\phi\ll 1$, the parameter $n$ determines the form of the potential minimum, and therefore the dynamics during reheating. T-model inflation can be easily accommodated within the no-scale supergravity models, where the parameter $\alpha$ is related to the curvature of the internal K\"ahler manifold~\cite{Ellis:2019bmm,Garcia:2020eof}. 

In the slow-roll approximation (\ref{eq:Nstarslowroll}), the value of the inflaton field at the horizon exit of the pivot scale $k_{\star}$ can be estimated as
\begin{equation}
\phi_{\star} ~\simeq~ \sqrt{\frac{3\alpha}{2}}  \cosh^{-1}\left[ \frac{4nN_{\star}}{3\alpha} + \cosh\left(\sqrt{\frac{2}{3\alpha}}\phi_{\rm end}\right)\right]\,,
\end{equation}
where
\begin{equation}
\phi_{\rm end} ~\simeq~ \sqrt{6\alpha } \coth ^{-1}\left(\frac{\sqrt{24 \alpha +4 \sqrt{3 \alpha  \left(12 \alpha +16 n^2-1\right)}+16 n^2-1}}{4 n-1}\right)\,, 
\end{equation}
is the inflaton value at the end of accelerated expansion. A simple computation using Eqs.~\eqref{eq:ns} and \eqref{eq:r} reveals that the CMB observables can be approximated as
\begin{subequations}
\label{eqs:nsTrT}
\begin{align} 
\label{eq:nsT}
n_s &\simeq~ 1-\frac{8 n^2 (3 \alpha +4 N_{\star})}{16 n^2 N_{\star}^2-9 \alpha ^2} ~\simeq~ 1-\frac{2}{N_{\star}}\,,\\ 
\label{eq:rT}
r &\simeq~ \frac{192 \alpha  n^2}{16 n^2 N_{\star}^2-9 \alpha ^2} ~\simeq~ \frac{12\alpha}{N_{\star}^2}\,,
\end{align}
\end{subequations}
where the second equality holds for $\alpha\lesssim \mathcal{O}(1)$. These simple expressions for the scalar tilt and the tensor-to-scalar ratio are a particular feature of attractor-like models, which also include Higgs inflation~\cite{Bezrukov:2007ep}.

For our present study we consider two particular realizations of the T-model, which correspond to the choices $n=1,2$. We scan over a wide range of values $10^{-2}<\alpha<10^2$, and over the nominal range $50<N_{\star}<60$. The results for the $\chi^2$ analysis for both scenarios are shown as the rightmost bottom panels in Fig.~\ref{fig:anisotropies_plot}. For both values of $n$, the combined {\em Planck}+BK analysis rules out $\alpha \lesssim 7\times 10^{-2}$ and $\alpha \gtrsim 10$ at $2\sigma$. To be within the $1\sigma$ preferred region, we must have $0.1 \lesssim \alpha\lesssim 8$. In addition, large values of $N_{\star}$ are preferred, with $N_{\star}\gtrsim52$ for $n=1$ and $N_{\star}\gtrsim 53$ for $n=2$. As expected from Eqs.~\eqref{eqs:nsTrT}, the constraint on $n_s$ is determined by $N_{\star}$, while the tensor constraint is mostly determined by $\alpha$. 

\begin{table}[t!]
	\centering
	\begin{tabular}{l|ll|lll|l}
		\toprule
		Point & $\alpha$ & $N_{\star}$ & $n_s$ & $r$ & $\frac{\mathrm{d}\,n_s}{\mathrm{d\,ln}\,k}$ & $\chi^2$ \\ \midrule
		Point A $(n=1)$ & $4$ & $53$ & $0.963$ & $0.0151$ & $-0.000288$ & $1.1$ \\
		Point B $(n=1)$ & $0.1$ & $56$ & $0.964$ & $0.0004$ & $-0.0100$ & $1.3$ \\\midrule
		Point A $(n=2)$ & $4$ & $56$ & $0.964$ & $0.0141$ & $-0.000262$ & $0.7$ \\
		Point B $(n=2)$ & $0.1$ & $56$ & $0.964$ & $0.0004$ & $-0.0100$ & $1.3$ \\\bottomrule
	\end{tabular}
	\caption{Sample points for T-model inflation  
	together with their CMB predictions. See also Fig.~\ref{fig:anisotropies_plot}.
	\label{tab:Tmodel}}
\end{table}

In the parameter space presented in Fig.~\ref{fig:anisotropies_plot} we focus on four particular points, listed in Table~\ref{tab:Tmodel} and shown as `$+$' and `$\times$' in the two bottom right panels of the figure. These points are selected not only because they lie within the 68\% CL {\em Plack} contours for $r$, $n_s$ and $n_s$ running, but also because they are consistent with the constraints imposed on $N_{\star}$ due to the duration of reheating. For $n=1$, the coherent oscillation of the inflaton mimics a matter dominated universe, $w\simeq 0$, and the last term in (\ref{eq:Nstar}) depends on the inflaton decay rate. Parametrizing it in terms of an effective Yukawa coupling $y$,
\begin{equation}
\Gamma_{\phi} \;=\; \frac{y^2}{8\pi}m_{\phi}\,,
\end{equation}
where $m_{\phi}=\Lambda^2/\sqrt{3\alpha}$, the perturbativity bound $y<1$ results in $N_{\star}\lesssim 55$ ($N_{\star}\lesssim 56$) for $\alpha=0.1$ ($\alpha=4$)~\cite{Ellis:2021kad}. On the other hand, for $n=2$, the oscillation of $\phi$ mimics a radiation dominated universe, $w\simeq 1/3$, and the number of $e$-folds become insensitive to the reheating epoch. For $\alpha=0.1$ ($\alpha=4$), one gets $N_{\star}\simeq 55$ ($N_{\star}\simeq 56$)~\cite{Garcia:2020eof,Garcia:2020wiy}.

\begin{figure}[!h!]
	\includegraphics[scale=0.26]{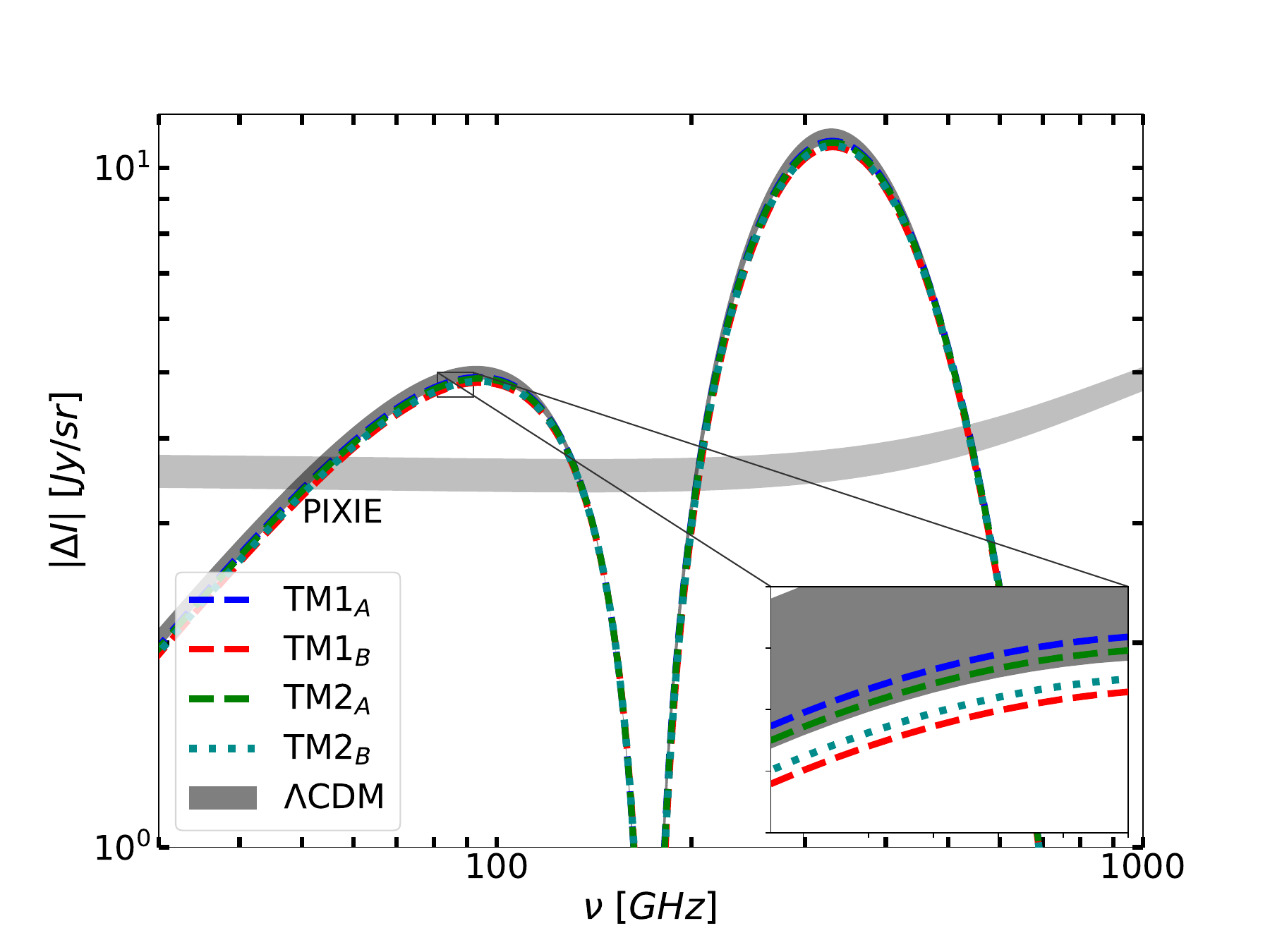}
	\includegraphics[scale=0.26]{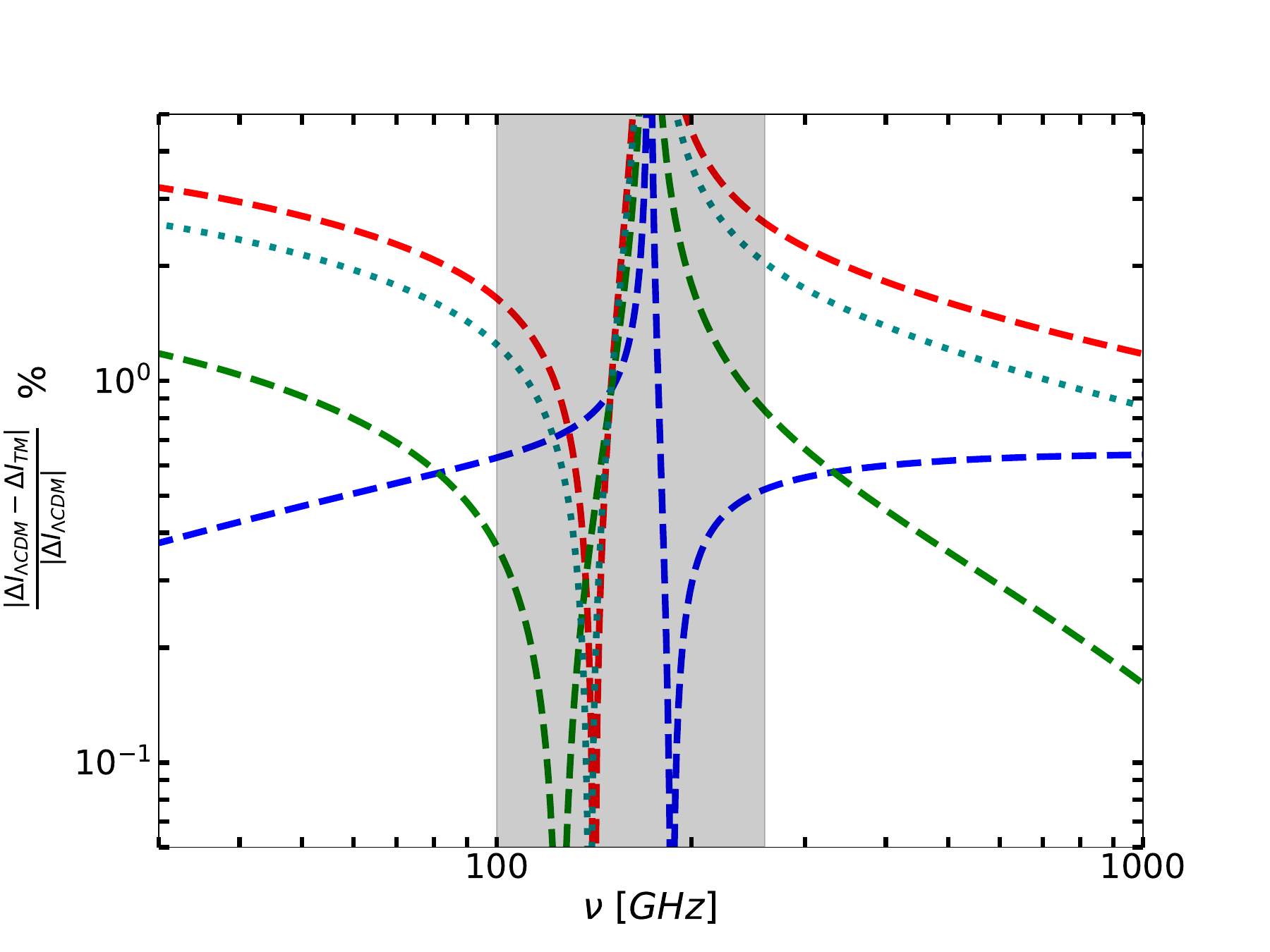}
	\caption{Left: predicted $\Delta I$ for the points in Table~\ref{tab:Tmodel} in the T-model, contrasted against the $\Lambda$CDM prediction dark gray region and the PIXIE projected sensitivity (gray region). The dashed curves correspond to the points A (TM1$_{A}$) and B (TM1$_{B}$) for $n=1$, and A (TM2$_{A}$) and B (TM2$_{B}$) for $n=2$. Right: difference between the T-model prediction $\Delta I_\mathrm{TM}$ and $\Delta I_{\Lambda\mathrm{CDM}}$. For frequencies outside the range $100\,\mathrm{GHz}\lesssim\nu\lesssim250\,\mathrm{GHz}$ the difference w.r.t.\ the fiducial signal is about 0.5\%-3\%.
	\label{fig:tmodel}}
\end{figure}

The signals of SDs predicted by the four different parameter choices in Table~\ref{tab:Tmodel} are shown as the dashed curves in the left panel of Fig.~\ref{fig:tmodel}, labeled as TM1$_{A}$ (point A)  and TM1$_{B}$ (point B) for $n=1$, and similarly TM2$_{A}$ and TM2$_{B}$ 
for $n=2$. For contrast, we also present the $\Lambda$CDM prediction and the sensitivity of the PIXIE experiment. As for the models discussed previously, the right panel of this figure depicts the percentage difference between the T-model signal $\Delta I_\mathrm{TM}$ and the $\Lambda$CDM result. In this case, the difference reaches values up to about 0.5\%-3\% approximately in the physically relevant region, $\nu\lesssim100$\,GHz and $\nu\gtrsim250$\,GHz.

We abstain from repeating the previous analysis for the CMB observables and the spectral distortions for the $\alpha$-Starobinsky models. Instead, we will discuss in detail the embedding of the Starobinsky model in a particular string-inspired multifield model in the following section.

\section{Multifield inflation}
\label{sec:multifield}
In the previous section, we studied the CMB compatibility and the SDs for a diverse set of single-field inflationary models. We now turn to the characterization of the predictions of three cases of multifield inflation based on the dynamics of two fields. Let us first study the power spectra for a generic two-field inflationary model, in order to quantify the relevance of multifield dynamics on the CMB temperature distribution.

For generality, we consider that the scalar fields that drive inflation are defined on a non-flat field manifold. Denoting them by $\phi_i$, with $i=1,2$, the action can then be written as
\begin{equation}
S \;=\; \int d^4\boldsymbol{x}\,\sqrt{-g}\left[-\frac{R}{2} + \frac{1}{2}G_{ij}\partial_{\mu}\phi^i\partial^{\mu}\phi^j - V(\boldsymbol{\phi})\right]\,,
\end{equation}
where $G_{ij}(\boldsymbol{\phi})$ denotes the metric in field space; for canonically normalized fields $G_{ij}=\delta_{ij}$. In this case, the background equations of motion can be written as
\begin{subequations}
\begin{align}
\ddot{\phi}^i + \Gamma^i_{jk}\dot{\phi}^j\dot{\phi}^k + 3H\dot{\phi}^i + G^{ij}V_{,j} &=~ 0\,, \\
\frac{1}{2}G_{ij}\dot{\phi}^i\dot{\phi}^j + V \;&=~ 3H^2\,,
\end{align}
\end{subequations}
where $V_{,j}$ denotes the derivative of $V$ w.r.t.\ $\phi^j$. Further,
$G^{ij}$ is the inverse field metric, and $\Gamma^i_{jk}$ denote the corresponding connection coefficients.

In the two-field case, two independent gauge-invariant scalar perturbations can be identified. They are given by the generalization of the Mukhanov-Sasaki variable, and we denote them by $Q^i$. In the spatially flat gauge, they are just $Q^i=\delta\phi^i$. It is nevertheless convenient to introduce a kinematical basis over the background trajectory, for which the components of the perturbations can be directly related to the curvature and isocurvature fluctuations~\cite{Gordon:2000hv,Tsujikawa:2002qx,DiMarco:2005nq,Byrnes:2006fr,Lalak:2007vi,Peterson:2011yt,Ellis:2014opa}. This basis corresponds to the instantaneous parallel and orthogonal directions to the background trajectory. Introducing the speed in field space
\begin{equation}
\dot{\sigma}^2 ~=~ G_{ij}\dot{\phi}^i\dot{\phi}^j \,,
\end{equation}
the parallel and orthogonal directions are determined by the orthonormal basis defined by
\begin{equation}
e_{\sigma}^i ~:=~ e_{\parallel}^i ~=~ \frac{\dot{\phi}^i}{\dot{\sigma}}\,, 
\qquad e_s^i ~:=~ e_{\perp}^i ~=~ \widetilde{G}^i_j\frac{\dot{\phi}^j}{\dot{\sigma}}\,,
\end{equation}
where $\widetilde{G}^i_j:=\epsilon^{ik}G_{kj}/\sqrt{G}$ with $\epsilon^{12}=1$. 
In this basis, the adiabatic and isocurvature perturbations are given respectively by
\begin{equation}
Q_{\sigma} ~=~ G_{ij}e^{i}_{\sigma}Q^j
\qquad \text{and}\qquad
Q_s ~=~ G_{ij}e^{i}_{s}Q^j\,,
\end{equation}
and satisfy the equations of motion
\begin{subequations}
\label{eqs:QsigmaQs}
\begin{align}
\label{Qsig_eq}
\ddot{Q}_{\sigma} + 3H\dot{Q}_{\sigma} + 2\frac{V_s}{\dot{\sigma}}\dot{Q}_s+ \left(\frac{k^2}{a^2}+C_{\sigma\sigma}\right) Q_{\sigma} + C_{\sigma s}Q_s &=~ 0\,,\\ 
\label{Qs_eq}
\ddot{Q}_s + 3H\dot{Q}_s - 2\frac{V_s}{\dot{\sigma}}\dot{Q}_{\sigma}+ \left(\frac{k^2}{a^2}+C_{ss}\right)Q_s + C_{s\sigma}Q_{\sigma} & =~ 0 \,.
\end{align}
\end{subequations}
The background-dependent coefficients are
\begin{subequations}
\begin{align}
C_{\sigma\sigma} & = \; V_{\sigma\sigma}-\left(\frac{V_s}{\dot{\sigma}}\right)^2 + \frac{2\dot{\sigma}}{H}V_{\sigma}+ 3\dot{\sigma}^2 - \frac{\dot{\sigma}^4}{2H^2}+ \Gamma^\ell_{ik}G_{\ell j}\dot{\phi}^i\dot{\phi}^j\dot{\phi}^k\frac{V_{\sigma}}{\dot{\sigma}^3} + \epsilon_{i\ell}\Gamma^\ell_{jk}\dot{\phi}^i\dot{\phi}^j\dot{\phi}^k\frac{V_s}{\dot{\sigma}^3}\,,\\
\notag C_{\sigma s} & = \; 6H\frac{V_s}{\dot{\sigma}}+2\frac{V_s V_{\sigma}}{\dot{\sigma}^2} + 2V_{\sigma s} + \frac{\dot{\sigma}V_s}{H} -2\widetilde{G}^\ell_iG_{mk}\Gamma^m_{\ell j}\dot{\phi}^i\dot{\phi}^j\dot{\phi}^k\frac{V_{\sigma}}{\dot{\sigma}^3}\\
&\quad  - 2\widetilde{G}^m_i\widetilde{G}^\ell_jG_{n\ell}\Gamma^n_{km}\dot{\phi}^i\dot{\phi}^j\dot{\phi}^k\frac{V_s}{\dot{\sigma}^3}\,,\\
C_{s\sigma} &= \; - 6H\frac{V_s}{\dot{\sigma}} - 2\frac{V_s V_{\sigma}}{\dot{\sigma}^2} + \frac{\dot{\sigma}V_s}{H}\,,\\
\notag C_{ss} &= \; V_{ss} - \left(\frac{V_s}{\dot{\sigma}}\right)^2 - \widetilde{G}^\ell_j\widetilde{G}^m_kG_{in}\Gamma^n_{\ell m}\dot{\phi}^i\dot{\phi}^j\dot{\phi}^k\frac{V_{\sigma}}{\dot{\sigma}^3} + \widetilde{G}^\ell_kG_{jm}\Gamma^m_{i\ell}\dot{\phi}^i\dot{\phi}^j\dot{\phi}^k \frac{V_s}{\dot{\sigma}^3}\\
&\quad - \frac{1}{2}\left(\widetilde{G}^k_i\widetilde{G}^{mj}G_{m\ell}\Gamma^\ell_{kj} + \widetilde{G}^k_i\Gamma^\ell_{k\ell}\right)\dot{\phi}^i\frac{V_s}{\dot{\sigma}} + \frac{1}{2}R\dot{\sigma}^2\,,
\end{align}
\end{subequations}
where $R$ denotes the curvature scalar, $\epsilon_{12}=\sqrt{G}\epsilon^{12}$, $\widetilde{G}^{ij}=G^{-1}\epsilon^{ik}\epsilon^{j\ell}G_{k\ell}$, and
\begin{equation}
V_{\sigma} ~=~ e_{\sigma}^iV_{,i}\,,\quad
V_{s} ~=~ e_{s}^iV_{,i}\,,\quad 
V_{\sigma \sigma} ~=~ e_{\sigma}^i e_{\sigma}^jV_{,ij}\,,\quad 
V_{\sigma s} ~=~ e_{\sigma}^i e_{s}^j V_{,ij}\,,\quad
V_{ss} ~=~ e_{s}^ie_{s}^j V_{,ij} \,.
\end{equation}
Upon solution of the equations of motion, the curvature and isocurvature power spectra can be computed as
\begin{subequations}
\begin{align}
\langle \mathcal{R}_k \mathcal{R}^*_{k'}\rangle \;&=\; \frac{2\pi^2}{k^3}\mathcal{P}_{\mathcal{R}}(k)\delta(k-k')\,,\\
\langle \mathcal{S}_k \mathcal{S}^*_{k'}\rangle \;&=\; \frac{2\pi^2}{k^3}\mathcal{P}_{\mathcal{S}}(k)\delta(k-k')\,,
\end{align}
\end{subequations}
with
\begin{equation}
\mathcal{R} ~=~ \frac{H}{\dot{\sigma}}Q_{\sigma}
\qquad\text{and}\qquad 
\mathcal{S} ~=~ \frac{H}{\dot{\sigma}}Q_s\,.
\end{equation}
If the motion in field space is geodesic, an analogue of the slow-roll approximation can be introduced, allowing for an analytical estimate of the CMB observables~\cite{Lalak:2007vi}. However, when this is not the case, the numerical solution of the system~\eqref{eqs:QsigmaQs} is necessary to determine the power spectra, and the corresponding amplitudes and tilts. For the three examples that we discuss below we follow this numerical approach.

\subsection{Quadratic multifield inflation}
\label{quadmulti}

Let us now explore a basic multifield scenario, consisting of two massive non-interacting fields defined over a flat manifold. Specifically, $G_{ij}=\delta_{ij}$ and
\begin{equation}
\label{eq:multiquad}
V(\boldsymbol{\phi}) ~=~ \frac{1}{2}m_1^2\phi_1^2 + \frac{1}{2}m_2^2\phi_2^2\,.
\end{equation}
\begin{table}[t!]
	\centering
	\begin{tabular}{l|rrr|ll}
		\toprule
		Point & $m_1$ ($\times 10^{-6}$) & $\phi_{1,0}$ & $\phi_{2,0}$ & $n_s$ & $r$  \\ \midrule
		QM$_1$ & $6.43$ & $17.82$ & $0.00$ & $0.963$ & $0.144$  \\
		QM$_2$ & $6.39$ & $17.71$ & $2.02$ & $0.962$ & $0.144$  \\ 
		QM$_3$ & $6.13$ & $17.04$ & $5.19$ & $0.959$ & $0.144$  \\
		QM$_4$ & $5.52$ & $15.67$ & $8.47$ & $0.953$ & $0.144$  \\ 
		QM$_5$ & $4.73$ & $13.67$ & $11.41$ & $0.953$ & $0.144$  \\ 
		QM$_6$ & $4.15$ & $11.53$ & $13.57$ & $0.956$ & $0.144$  \\ 
		QM$_7$ & $3.73$ & $9.03$ & $15.34$ & $0.959$ & $0.145$  \\ 
		QM$_8$ & $3.49$ & $6.84$ & $16.43$ & $0.961$ & $0.145$  \\ 
		QM$_9$ & $3.33$ & $4.23$ & $17.28$ & $0.962$ & $0.145$  \\ 
		QM$_{10}$ & $3.23$ & $1.83$ & $17.74$ & $0.963$ & $0.144$  \\ 
		QM$_{11}$ & $3.21$ & $0.00$ & $17.82$ & $0.963$ & $0.144$  \\ \bottomrule
	\end{tabular}
	\caption{Sample points for quadratic multifield inflation, together with their CMB predictions. See also Fig.~\ref{fig:PRquad}.
	\label{tab:QMI}}
\end{table} 
\begin{figure}[!t]
\centering
    \includegraphics[width=0.96\textwidth]{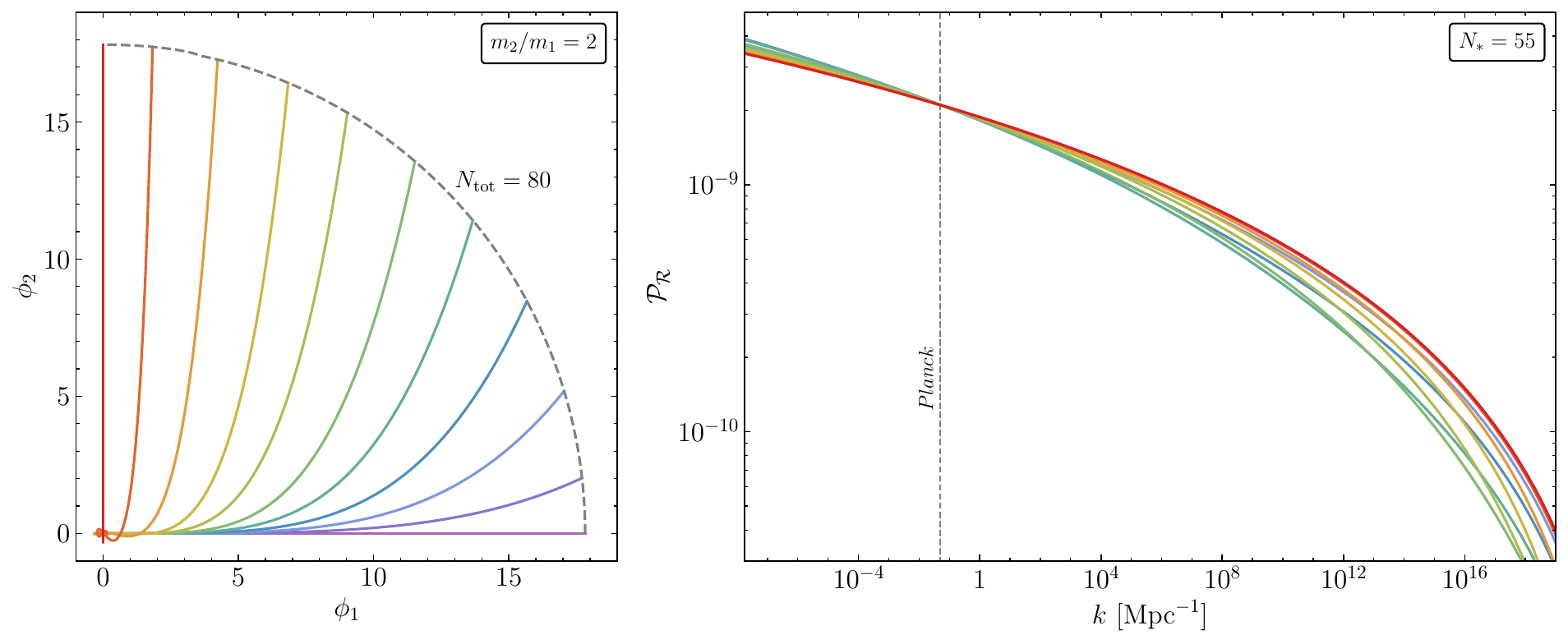}
    \caption{Left: field-space trajectories for the quadratic multifield model~\eqref{eq:multiquad}, with initial conditions chosen so that $N_{\rm tot}=80$. Right: curvature power spectra computed along the selected field trajectories, with the same color coding, assuming the pivot scale $k_{\star}=0.05\,{\rm Mpc}^{-1}$ leaves the horizon 55 $e$-folds before the end of inflation.
    \label{fig:PRquad}}
\end{figure}%
In order to have a comparable field displacement in both fields, while avoiding having trivial trajectories, we fix $m_2/m_1=2$. The left panel of Fig.~\ref{fig:PRquad} shows the selection of trajectories that we consider for our analysis. The corresponding initial conditions, Lagrangian parameters and CMB observables are shown in Table~\ref{tab:QMI}. Due to the extended freedom in initial conditions, we set $\dot{\phi}_i=0$, and choose field values for which we have a total of 80 $e$-folds of expansion during inflation. The set of points that satisfies this constraint is shown as the gray dashed line. Along each of these trajectories we solve the system~\eqref{eqs:QsigmaQs} in order to obtain the curvature (and isocurvature) power spectrum. For definiteness we assume 55 $e$-folds between the horizon exit of the {\em Planck} pivot scale and the end of inflation. The resulting curvature power spectra for each field trajectory is shown in the right panel of Fig.~\ref{fig:PRquad}, with the same color coding. Here we have chosen $m_1$ so that the {\em Planck} normalization on the power spectrum amplitude is enforced, as emphasized by the vertical gray line. For the considered initial conditions this requires $6.4\times 10^{-6}\lesssim m_1/M_P \lesssim 3.2\times 10^{-6}$. The isocurvature power spectrum rapidly decays outside the horizon in all cases, and can therefore be safely disregarded. 

It is worth noting how the bending of the background trajectory leaves an imprint on the curvature power spectrum. For the green curves, for which almost the complete trajectory is curved in field space, the resulting power spectrum is steeper at the {\em Planck} pivot scale. Accordingly, these spectra lead to the smallest values of $n_s\simeq0.953$, which lie outside of the marginalized 95\% CL region for the tilt. On the other hand, the tilt is closest to one, more precisely $n_s\simeq0.963$, for the two straight, purple and red trajectories, with initial conditions $\phi_{1,2}=0$. 

The absence of interaction between the two fields in this model implies that the sourcing of $\mathcal{R}$ modes from isocurvature modes is not large enough to leave a significant impact on the tensor-to-scalar ratio. Indeed, one can show that $r\simeq0.14$ for all cases, which is close to the single field prediction and far away from the current 2$\sigma$ CL contours, ruling out the model (\ref{eq:multiquad}) as a realistic inflationary scenario.

\begin{figure}[t!]
    \includegraphics[scale=0.26]{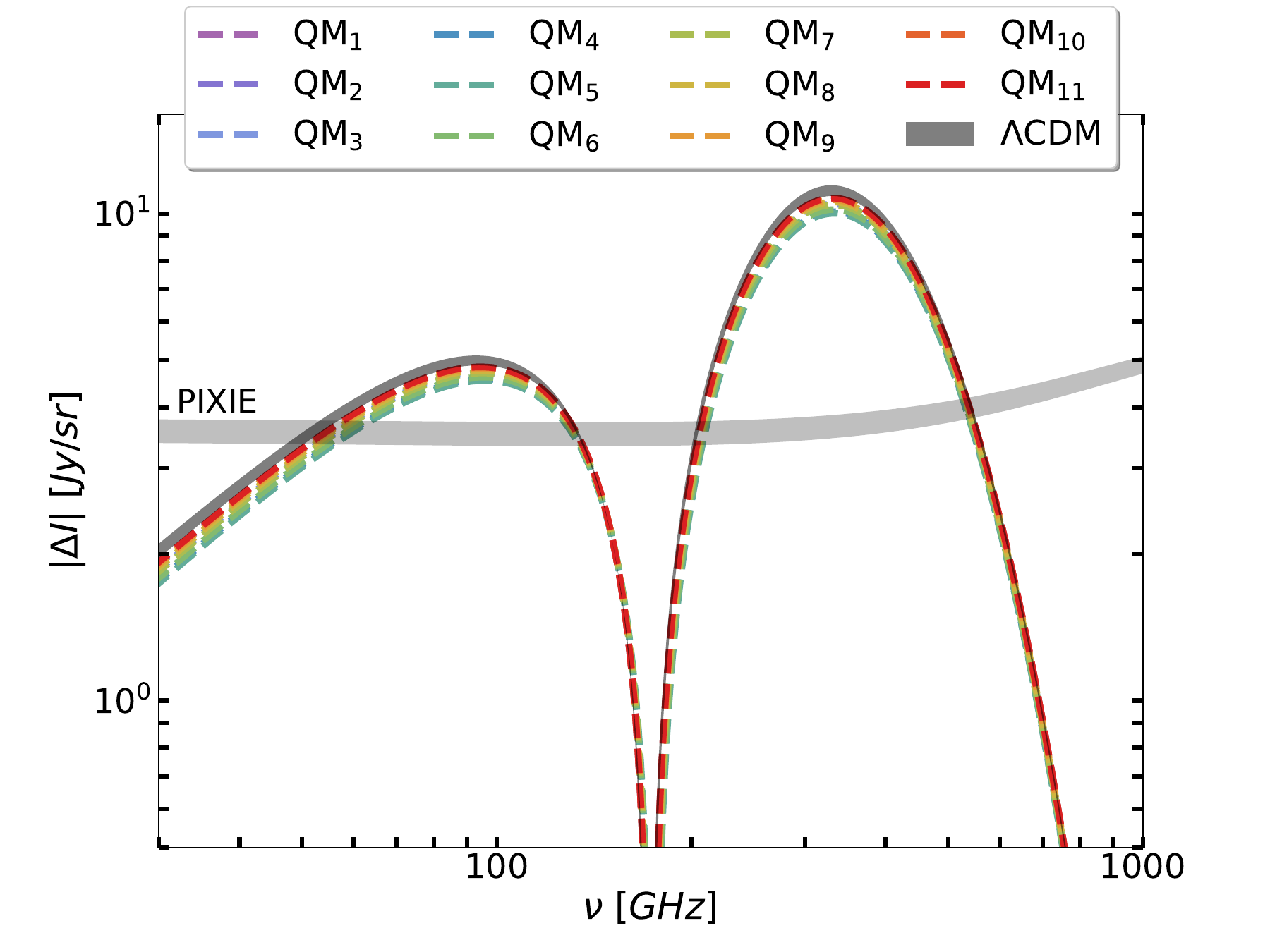}
    \includegraphics[scale=0.26]{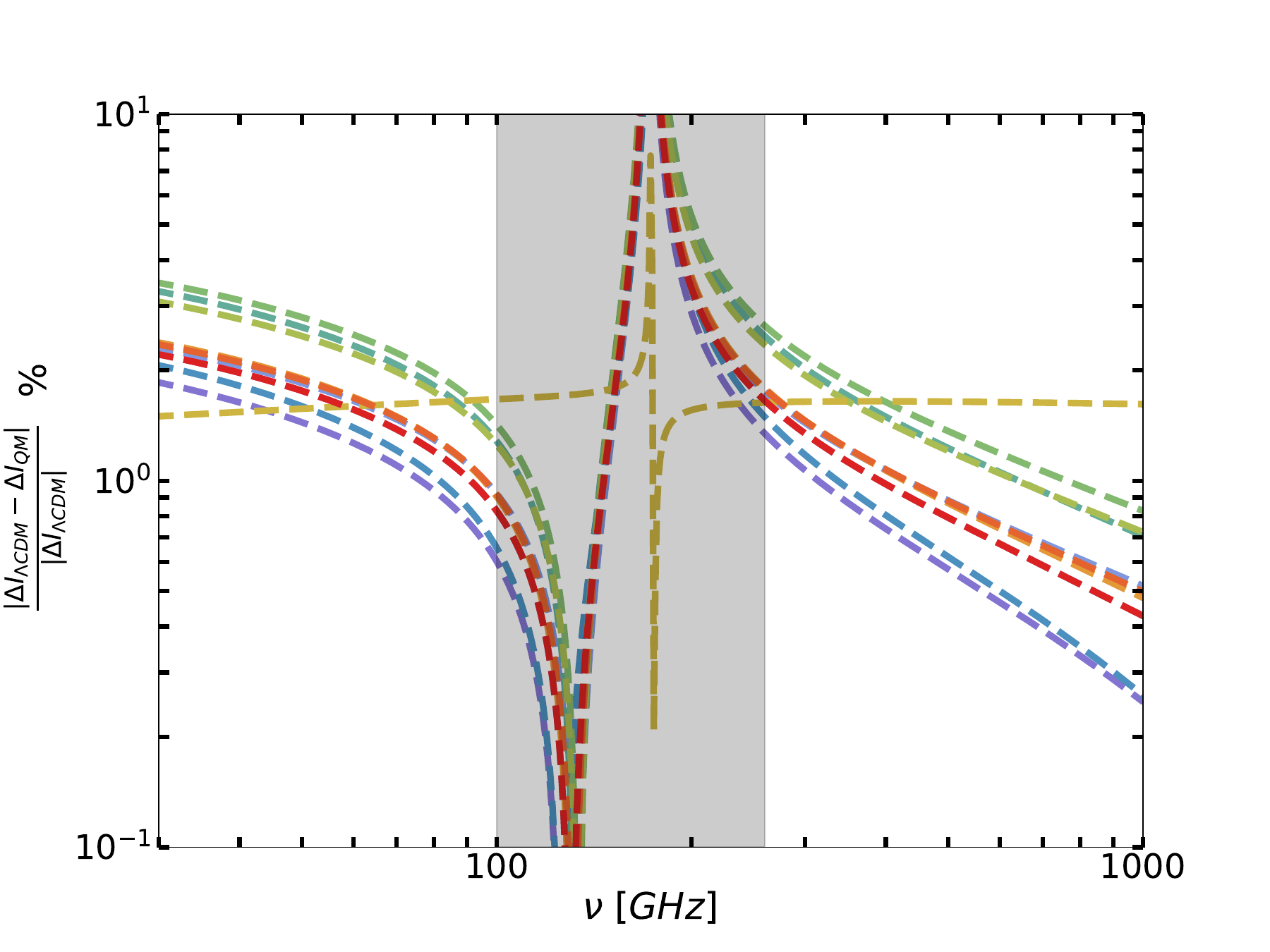}
    \caption{Left: predicted $\Delta I$ for the data selection from Fig.~\ref{fig:PRquad} in the quadratic multifield (QM) model~\eqref{eq:multiquad}, contrasted against the $\Lambda$CDM prediction dark gray region and the projected PIXIE sensitivity (gray region). The curves share the same color coding as those in Fig.~\ref{fig:PRquad}, and are labeled as $QM_{N}$ with $N=1,2,...,11$. Right: percentage difference between $\Delta I_\mathrm{QM}$ and $\Delta I_{\Lambda\mathrm{CDM}}$. For frequencies outside the range $100\,\mathrm{GHz}\lesssim\nu\lesssim250\,\mathrm{GHz}$ the difference w.r.t.\ the fiducial signal is about 2\%-3\%.
    \label{fig:sdmultifield}}
\end{figure}

The signals of SDs predicted for the quadratic multifield model by the eleven different parameter choices from Fig.~\ref{fig:PRquad} are shown as the dashed curves in the left panel of Fig.~\ref{fig:sdmultifield}, labeled as $QM_{N}$ with $N=1,2,...,11$. Similarly to the single-field models discussed previously, the right panel of this figure depicts the percentage difference between the quadratic multifield model signal $\Delta I_\mathrm{QM}$ and the $\Lambda$CDM result. In this case, the difference reaches values up to about 2\%-3\% for the frequencies of interest.

\subsection{The EGNO model}

We now turn our focus to a more interesting model, for which the interactions between the two fields lead to a non-trivial manifold structure, and therefore, to a non-trivial evaluation of the inflationary observables. This model is defined in a no-scale supergravity framework, inspired by orbifold compactifications of strings~\cite{Bailin:1999nk,Ramos-Sanchez:2008nwx,Vaudrevange:2008sm} in which matter fields have non-vanishing modular weights~\cite{Dixon:1989fj,Ibanez:1992hc,Araki:2008ek,Antusch:2011ei,Olguin-Trejo:2017zav,Baur:2022hma}. Concretely, it is assumed that the ratios of the (three) orbifold K\"ahler moduli are fixed at a high scale, so that the K\"ahler potential can be written as
\begin{equation}
K ~=~ -3\ln(T+\overline{T}) + \frac{|\varphi|^2}{(T+\overline{T})^3}\,,
\end{equation}
with $T$ the (six-dimensional overall) volume modulus and $\varphi$ a singlet matter field with modular weight $-3$. The superpotential is chosen as
\begin{equation}
W ~=~ \sqrt{3}m\varphi\left(T-\frac{1}{2}\right)\,.
\end{equation}
This model is known as the EGNO model~\cite{Ellis:2014opa,Ellis:2014gxa,Aragam:2021scu}, and possesses two particular properties. The first is that this construction does not require any additional moduli stabilization to drive inflation~\cite{Ellis:1984bs,Ellis:2013nxa}. The inflationary solution corresponds to taking $\varphi=0$, a condition dynamically self-enforced, and identifying $T$ with the (complex) inflaton field. In terms of its canonically normalized real and imaginary components, 
\begin{equation}
T ~=~ \frac{1}{2}\left( e^{-\sqrt{\frac{2}{3}}\phi_1} + i \sqrt{\frac{2}{3}}\phi_2\right)\,,
\end{equation}
the effective Lagrangian takes the form
\begin{equation}\label{eq:egno}
\mathcal{L} \;=\; \frac{1}{2}\partial_{\mu}\phi_1\partial^{\mu}\phi_1 + \frac{1}{2}e^{2\sqrt{\frac{2}{3}}\phi_1}\partial_{\mu}\phi_2\partial^{\mu}\phi_2 - \frac{3}{4}m^2 \left( 1 - e^{-\sqrt{\frac{2}{3}}\phi_1}\right)^2 - \frac{1}{2}m^2 \phi_2^2\,.
\end{equation}
For purely real inflation ($\phi_2=0$), the Lagrangian reduces to the $R+R^2$ Starobinsky model~\cite{Starobinsky:1980te} in its scalaron form, a generic feature of no-scale supergravity constructions~\cite{Ellis:2013xoa,Ellis:2013nxa,Ellis:2018zya}. On the other hand, at $\phi_1=0$, $\mathcal{L}$ appears to reduce to the standard quadratic chaotic scenario. However, the presence of the non-trivial kinetic coupling, correlated with a non-Euclidean structure of the field manifold, drives the background dynamics away from the purely imaginary direction. The inflationary trajectories are non-geodesic, and the model cannot be simplified to an effective single-field framework. In other words, the kinetic coupling induces a coupling between the curvature and isocurvature perturbations, resulting in an enhancement of the curvature modes at super-horizon scales. This is the second main feature of the EGNO construction. 

\begin{table}[t!]
	\centering
	\begin{tabular}{l|rrr|ll}
		\toprule
		Point & $m$ ($\times 10^{-6}$) & $\phi_{1,0}$ & $\phi_{2,0}$ & $n_s$ & $r$  \\ \midrule
		EGNO$_1$ & $12.31$ & $5.76$ & $0.00$ & $0.965$ & $0.0036$  \\
		EGNO$_2$ & $9.27$ & $5.09$ & $1.06$ & $0.964$ & $0.0035$  \\ 
		EGNO$_3$ & $5.68$ & $3.94$ & $2.32$ & $0.964$ & $0.0034$  \\
		EGNO$_4$ & $4.04$ & $3.16$ & $3.45$ & $0.962$ & $0.0033$  \\ 
		EGNO$_5$ & $3.01$ & $2.55$ & $4.62$ & $0.960$ & $0.0032$  \\ 
		EGNO$_6$ & $2.43$ & $1.99$ & $6.00$ & $0.958$ & $0.0031$  \\ 
		EGNO$_7$ & $2.08$ & $1.62$ & $7.12$ & $0.956$ & $0.0030$  \\ 
		EGNO$_8$ & $1.77$ & $1.23$ & $8.53$ & $0.953$ & $0.0029$  \\ 
		EGNO$_9$ & $1.52$ & $0.86$ & $10.15$ & $0.950$ & $0.0028$  \\ 
		EGNO$_{10}$ & $1.32$ & $0.52$ & $11.94$ & $0.946$ & $0.0028$  \\ 
		EGNO$_{11}$ & $1.11$ & $0.00$ & $14.79$ & $0.941$ & $0.0027$  \\ \bottomrule
	\end{tabular}
	\caption{Sample points for the EGNO model, together with their CMB predictions. See also Fig.~\ref{fig:PRegno}.
	\label{tab:EGNO}}
\end{table} 

\begin{figure}[!t]
\centering
    \includegraphics[width=0.96\textwidth]{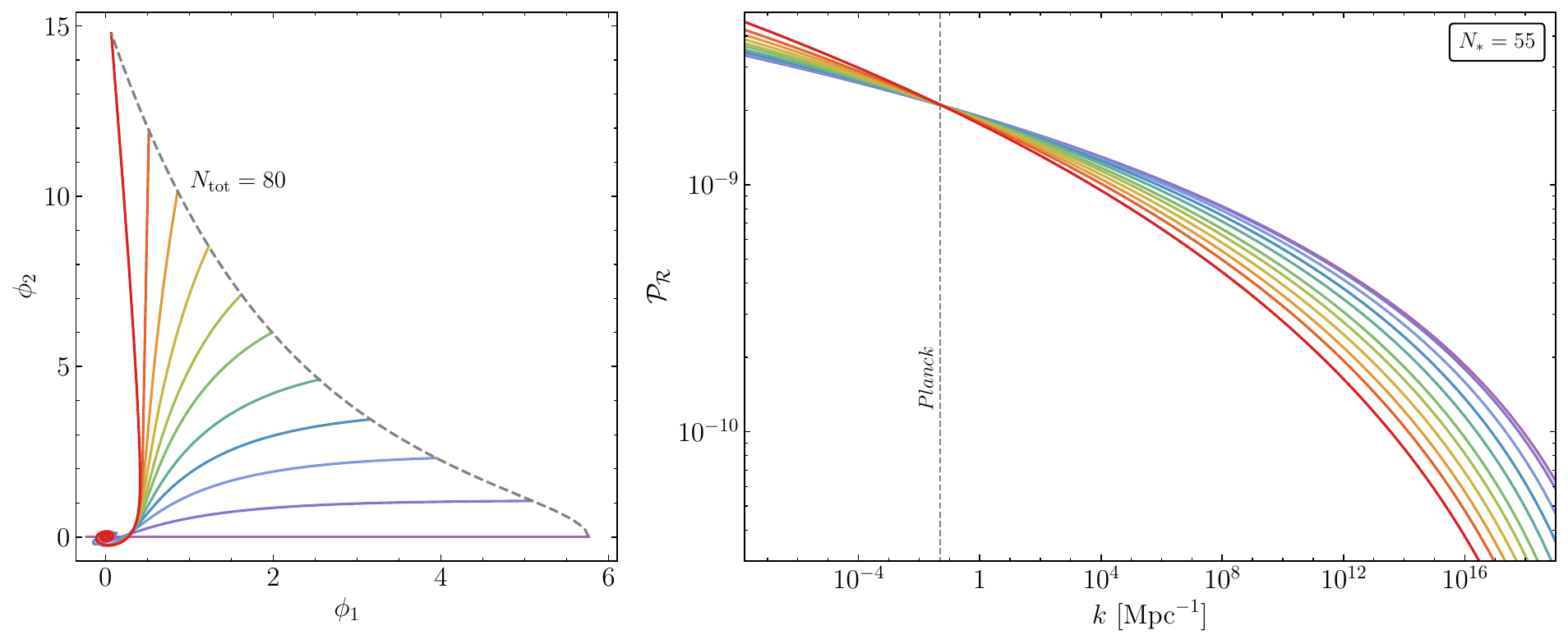}
    \caption{Left: field-space trajectories for the EGNO multifield model~\eqref{eq:egno}, with initial conditions chosen so that $N_{\rm tot}=80$. Right: the curvature power spectra computed along the selected field trajectories, with identical color coding, assuming the pivot scale $k_{\star}=0.05\,{\rm Mpc}^{-1}$ leaves the horizon 55 $e$-folds before the end of inflation.
    \label{fig:PRegno}}
\end{figure}

The left panel of Fig.~\ref{fig:PRegno} shows a selection of background trajectories in field space. For all of them, $N_{\rm tot}=80$ and we set $\dot{\phi}_1=\dot{\phi}_2=0$ as initial conditions. For the horizontal, purple trajectory, pure Starobinsky inflation is realized, as the kinetic coupling vanishes in this case. The isocurvature fluctuation is negligible, and the curvature fluctuation $\mathcal{R}$ is conserved outside the horizon. On the other hand, all the remaining trajectories are bent clockwise by the field-space curvature. This bending is enhanced at larger kinetic energies, and as a result the motion is made spiral by the end of inflation. The deviation away from geodesics leads to an enhancement of the curvature perturbation on superhorizon scales by the isocurvature fluctuation. This sourcing of $\mathcal{R}$ from $\mathcal{S}$ is maximal for the red trajectory with initial $\phi_1=0$. The corresponding initial conditions, Lagrangian parameters and CMB observables are shown in Table~\ref{tab:EGNO}.

\begin{figure}[t!]
    \centering
    \includegraphics[scale=0.65]{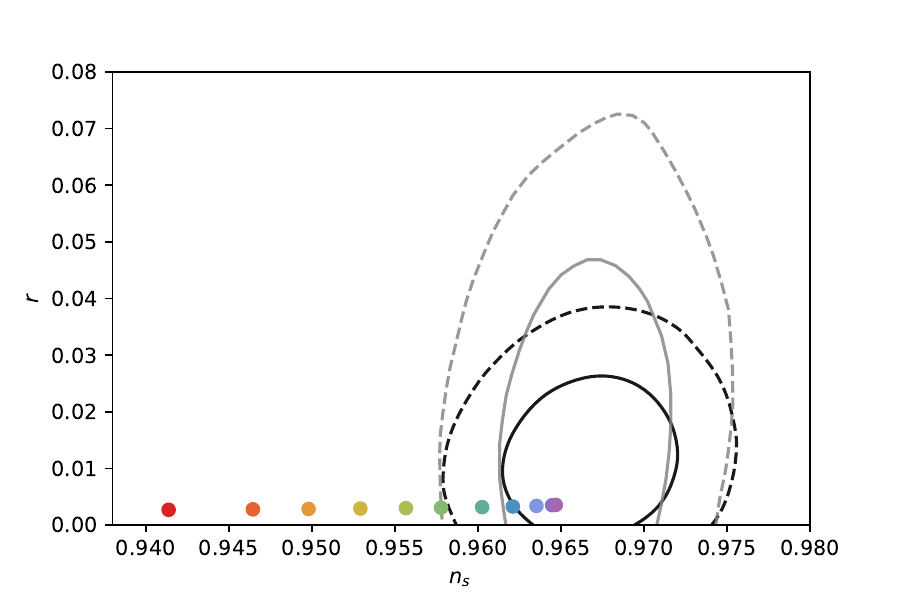}
    \caption{Inflationary CMB observables $n_s$ and $r$ for the multifield EGNO model. Each point is color coded with the field-space trajectories in Fig.~\ref{fig:PRegno}. The scalar tilt and the tensor-to-scalar ratio are evaluated numerically under the assumption that the {\em Planck} pivot scale $k_{\star}=0.05\,{\rm Mpc}^{-1}$ left the horizon 55 $e$-folds before the end of inflation.
    \label{fig:EGNO_ns_r}}
\end{figure}

The imprint of the multifield dynamics on the scalar power spectrum can be appreciated in the right panel of  Fig.~\ref{fig:PRegno}. By adjusting the mass parameter $m$, all resulting power spectra can be normalized at the {\it Planck} pivot scale.
The purple spectrum corresponds to the pure Starobinsky result, which has the smallest tilt. We note that the spectral tilt increases with increased geodesic deviation. Long wavelength modes experience a longer lasting sourcing from $\mathcal{S}$ since they exit the horizon sooner. The resulting spectrum for pure imaginary initial conditions, for example, is steeper than in the standard quadratic chaotic scenario, cf.~Fig.~\ref{fig:PRquad} (see~\cite{Ellis:2014opa} for further details). The corresponding values of the tilts $n_s$, with the same color coding, can be better appreciated in Fig.~\ref{fig:EGNO_ns_r}. There, we observe that only the three trajectories with initial condition for $\phi_2$ closest to zero lie within the 68\% {\em Planck}+BK18 CL contours. The next two (light blue and dark green) lie outside the 1$\sigma$ contour, but inside the 2$\sigma$ CL region. The rest of the trajectories, with larger initial $\phi_2$, yield scalar power spectra too steep to lie within the 95\% CL CMB preferred region.

\begin{figure}[t!]
    \includegraphics[scale=0.26]{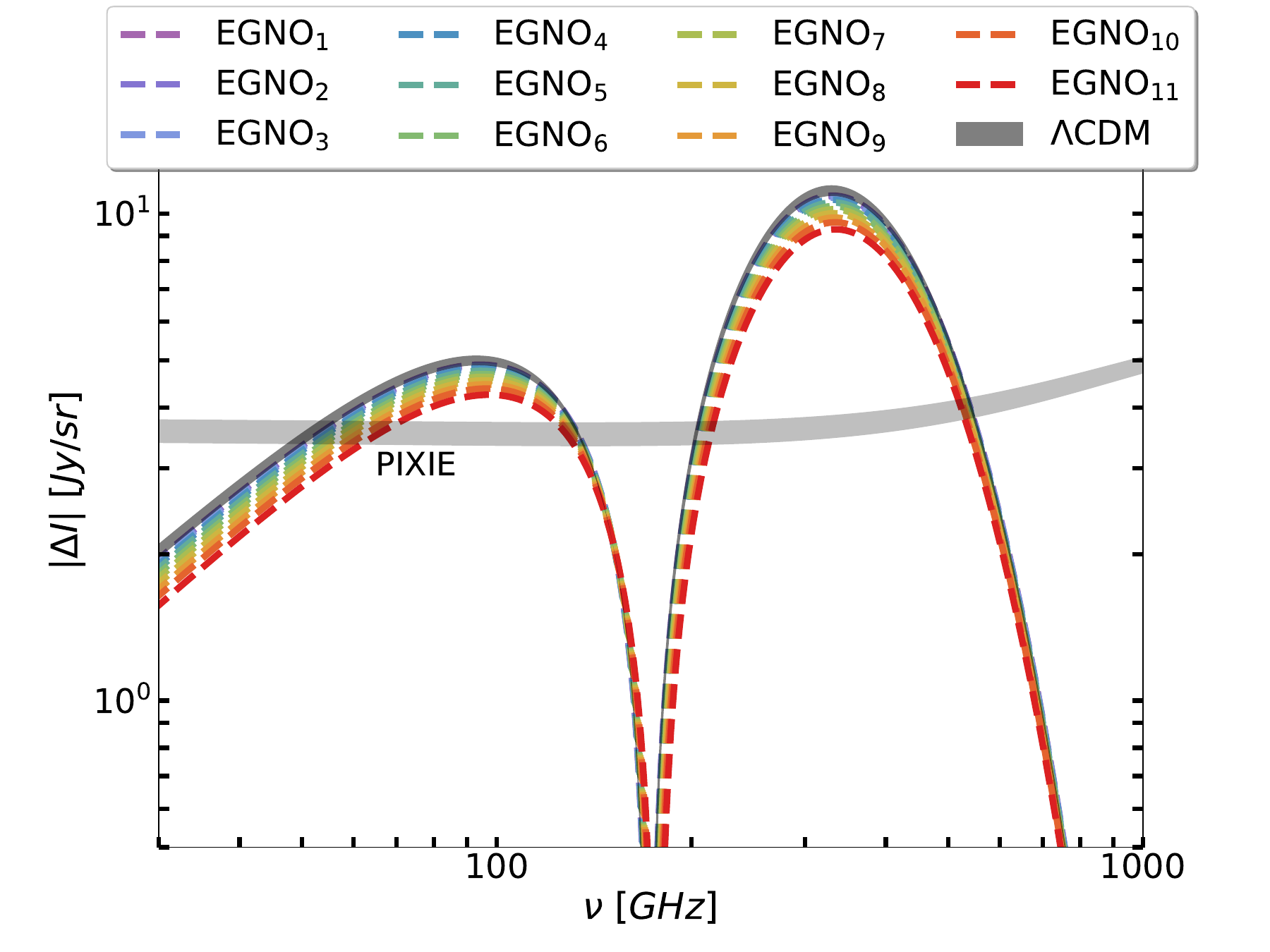}
    \includegraphics[scale=0.26]{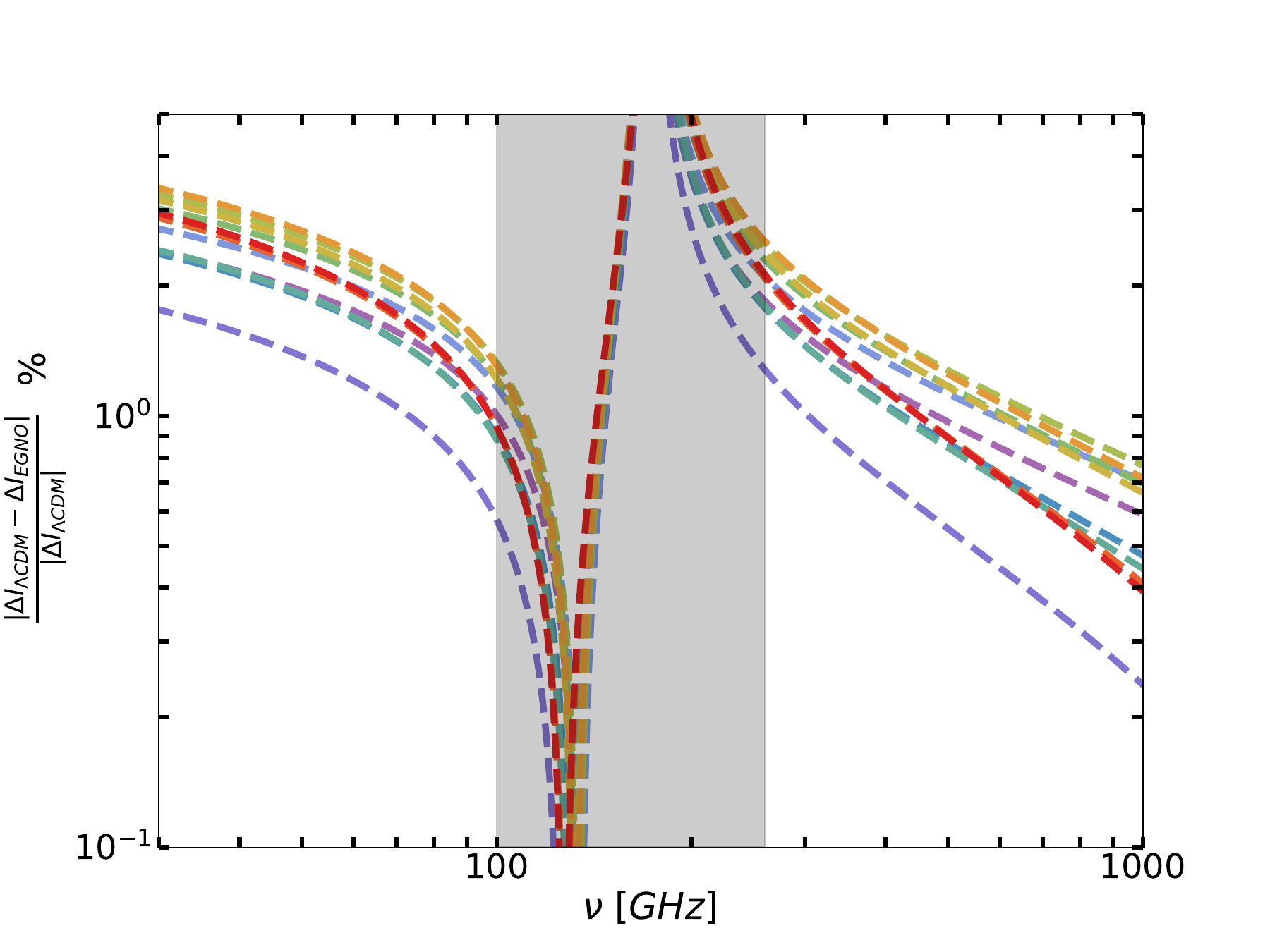}
    \caption{Left: $\Delta I$ for the EGNO model with the conditions given in Fig.~\ref{fig:PRegno}, contrasted against the $\Lambda$CDM prediction and PIXIE projected sensitivity, as in previous figures. The dashed curves share the color coding with the background solutions and spectra in Fig.~\ref{fig:PRegno}, and are labeled as $\mathrm{EGNO}_N$ with $N=1,2,...,11$. Right: percentage difference between the EGNO prediction $\Delta I_\mathrm{EGNO}$ and $\Delta I_{\Lambda\mathrm{CDM}}$. For frequencies outside the range $100\,\mathrm{GHz}\lesssim\nu\lesssim250\,\mathrm{GHz}$ the difference w.r.t.\ the fiducial signal is about 2\%-3\%.
    \label{fig:sdmultifieldegno}}
\end{figure}

The enhancement of the curvature spectrum in the presence of multifield effects not only alters the predictions for $n_s$, but leaves a stronger imprint on $r$. As the tensor spectrum is solely determined at linear order by the energy scale of inflation, its amplitude $A_T$ is impervious to multifield effects. The result is a suppression of $r$ relative to the corresponding single-field scenario. As Fig.~\ref{fig:EGNO_ns_r} shows, not only does the red ``quadratic'' trajectory of Fig.~\ref{fig:PRegno} give significantly lower values of $r$ ($r=0.0027$) than in the pure chaotic case explored in Section~\ref{quadmulti} ($r=0.14$), but yields a value of $r$ that is even smaller than in the pure Starobinsky purple case ($r=0.0035$). The result is that for all initial conditions the CMB constraint in $r$ is satisfied. 

The predicted SDs for the multifield EGNO model for the eleven different initial conditions in Fig.~\ref{fig:PRegno} are represented by the dashed curves in the left panel of Fig.~\ref{fig:sdmultifieldegno}, labeled as $EGNO_{N}$ with $N=1,2,...,11$, together with the $\Lambda$CDM prediction and the sensitivity of the PIXIE experiment. As the right panel shows, in this case the percentage difference between $\Delta I_\mathrm{EGNO}$ and the $\Lambda$CDM result lies in the range 2\%-3\% depending on the parameter choices, for $\nu\lesssim100$\,GHz and $\nu\gtrsim250$\,GHz.

\subsection{Hybrid attractors}
\label{sec:hybrid}

As studied in the previous case, large geodesic deviations alter the scalar and tensor spectra. In particular, when the deviation is originated from a kinetic-term coupling, the isocurvature sourcing is more prominent for long wavelength modes, as the geodesic deviation is active for the duration of inflation. The net effect is an enhancement of the $\mathcal{R}$ power spectrum for $k<k_{\star}$, increasing the red tilt of the spectrum, and thus yielding sub-$\Lambda$CDM SDs.

These results suggest that a sharp localized turn in the field manifold, capable of producing a blue-tilted spectrum, could also lead to a significant $\Delta I$ signal. These ingredients are naturally found in hybrid inflationary models~\cite{Linde:1991km,Linde:1993cn}. In this case, two fields are involved: $\phi$ is the inflaton field, responsible for the slow-roll phase that leads to the quasi-exponential expansion; and $\chi$, the so-called waterfall field, responsible for finishing inflation and for reheating the universe. The transition from $\phi$-dominated to $\chi$-dominated dynamics arises from the time-dependence of the effective mass of the waterfall field near the origin. The coupling between $\phi$ and $\chi$ induces an effective mass $m_{\chi}^2(\phi)$ that is large and positive above a certain critical value $\phi_c$. This stabilizes $\chi$ around zero, and allows for the slow-roll of $\phi$. When $\phi<\phi_c$, however, $m_{\chi}^2(\phi)<0$, i.e.\ the waterfall field $\chi$ becomes tachyonic. It is known that this situation leads to an 
exponential growth of its fluctuations that drives $\chi$ away from the origin towards the low-energy minimum $\chi=\chi_0$~\cite{Linde:1993cn}.

An undesired effect of the aforementioned growth of the $\chi$-fluctuations is that it enhances the curvature power spectrum at low scales, leading to $n_s>1$ for a wide range of wavenumbers. For this reason, hybrid inflation has not been part of the `mainstream' models since the measurement of a red tilt at CMB scales~\cite{Planck:2018jri}. Nevertheless, if the waterfall transition is sudden, and inflation can be sustained for a few $e$-folds afterwards, it is possible to restrict the enhancement of $\mathcal{P}_{\mathcal{R}}$ to scales lower than $k_{\star}$. If this is the case, CMB observables would be determined solely by the slow-roll phase of the dynamics. This scenario is reminiscent of inflection-point inflation models~\cite{Ivanov:1994pa,Hodges:1989dw,Destri:2007pv,Ballesteros:2015noa,Garcia-Bellido:2017mdw,Ballesteros:2017fsr}. 

Recent implementations of this idea are connected with the embedding in supergravity of hybrid inflation. A working model based on the addition of shift-symmetric terms to an otherwise canonical K\"ahler potential was introduced in~\cite{Spanos:2021hpk}. A more recent realization corresponds to a hybrid version of the $\alpha$-attractor models, which result from supergravity scenarios including stabilized axions~\cite{Kallosh:2022ggf,Braglia:2022phb}. In the canonically normalized basis, the potential can be written in the form
\begin{equation}\label{eq:hybridV}
V(\phi,\chi) \;=\; M^2\left[ \frac{(\chi^2-\chi_0^2)^2}{4\chi_0^2} + 3\alpha(m^2+g^2\chi^2)\tanh^2\left(\frac{\phi}{\sqrt{6}\alpha}\right) +d\chi \right]\,.
\end{equation}
Note that at $\chi=0$ the potential is simply
\begin{equation}
V(\phi,0) \;=\; M^2\left[ \frac{1}{4}\chi_0^2 + 3\alpha m^2\tanh^2\left(\frac{\phi}{\sqrt{6}\alpha}\right)  \right]\,,
\end{equation}
which, barring the constant term, takes the T-model form~\eqref{eq:VTmodel}, inheriting therefore its {\em Planck}-compatibility at CMB scales. This is ensured by the large mass of the waterfall field at large $\phi$ values,
\begin{equation}
m_{\chi}^2(\phi) \;=\; \partial_{\chi\chi}V(\phi,\chi)|_{\chi=0} \;=\; M^2\left[-1 +6\alpha g^2 \tanh^2\left(\frac{\phi}{\sqrt{6}\alpha}\right) \right]\,.
\end{equation}

\begin{figure}[!t]
\centering
	\includegraphics[width=0.66\textwidth]{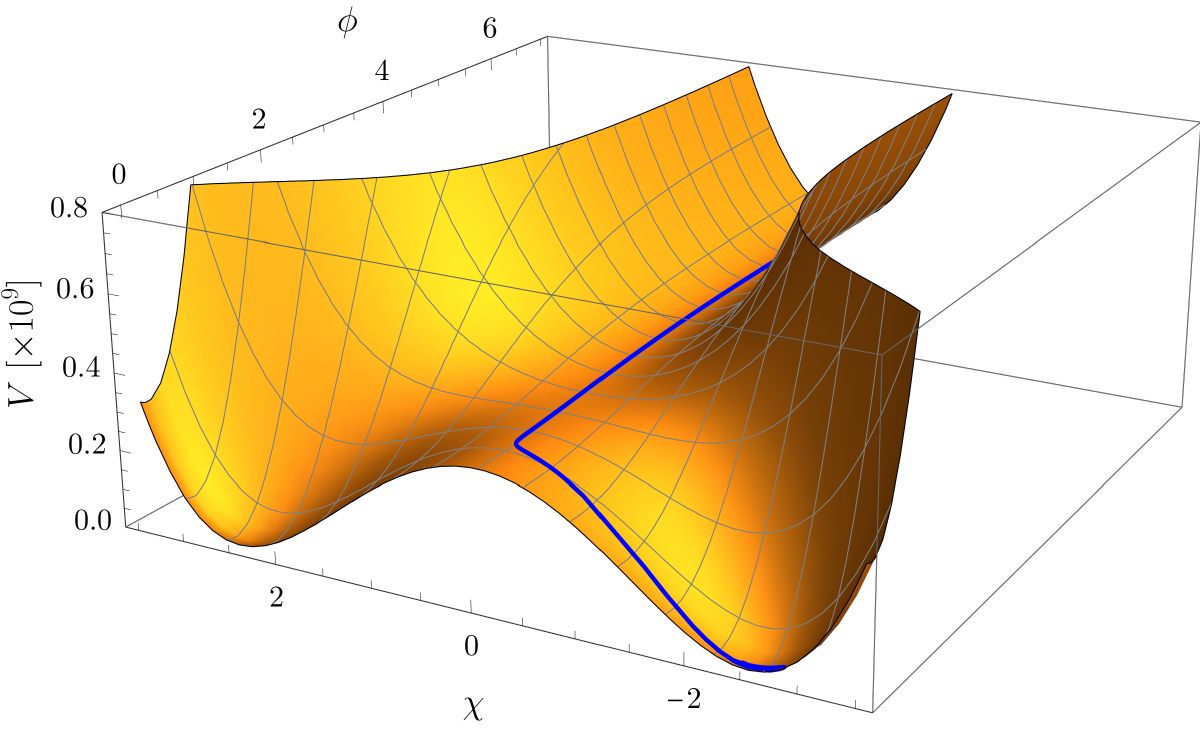}
	\caption{The scalar potential for the hybrid $\alpha$-attractor model~\eqref{eq:hybridV} with benchmark parameters~\eqref{eq:benchmark}. Shown as the blue curve is a particular field trajectory.
	\label{fig:HV}}
\end{figure}

\begin{table}[t!]
	\centering
	\begin{tabular}{l|cll|ll}
		\toprule
		Point              & $d$                 & $\chi_0$ & $g$   & $n_s$   & $r$ \\ \midrule
		H$_{d,{\rm A}}$    & $-5\times 10^{-6}$  & $2.5$    & $0.8$ & $0.962$ & $0.012$\\
		H$_{d,{\rm B}}$    & $-10^{-6}$          & $2.5$    & $0.8$ & $0.962$ & $0.011$\\
		H$_{d,{\rm C}}$    &  $-5\times 10^{-5}$ & $2.5$    & $0.8$ & $0.962$ & $0.010$\\
		H$_{\chi,{\rm B}}$ &  $-5\times 10^{-6}$ & $2.4$    & $0.8$ & $0.959$ & $0.012$ \\
		H$_{\chi,{\rm C}}$ &  $-5\times 10^{-6}$ & $2.3$    & $0.8$ & $0.956$ & $0.012$\\
		H$_{g,{\rm B}}$    &  $-5\times 10^{-6}$ & $2.5$    & $1.0$ & $0.964$ & $0.015$\\
		H$_{g,{\rm C}}$    &  $-5\times 10^{-6}$ & $2.5$    & $1.2$ & $0.967$ & $0.016$\\\bottomrule
	\end{tabular}
	\caption{Sample points for the hybrid attractor model~\eqref{eq:hybridV}, and their associated predictions for the CMB observables. The parameters $M$, $\alpha$ and $m$ are fixed to the benchmark values~\eqref{eq:benchmark}.
        \label{tab:hybrid}}
\end{table}

\begin{figure}[!t]
\centering
	\includegraphics[width=\textwidth]{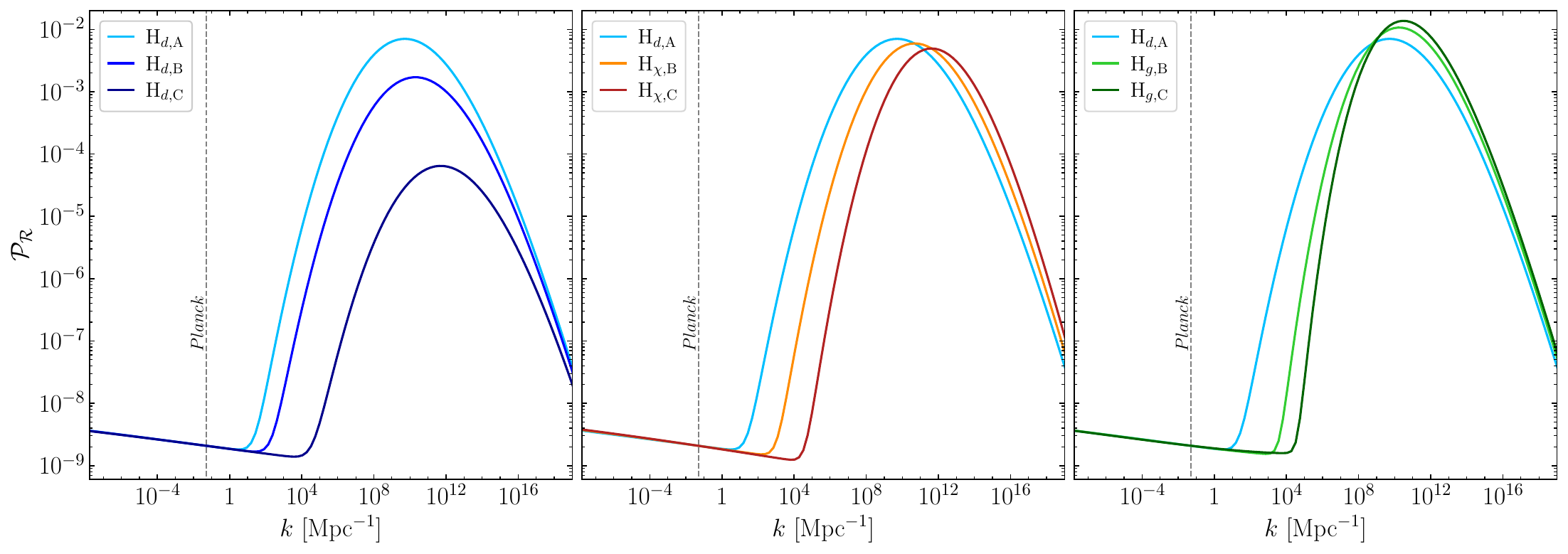}
	\caption{Curvature power spectra for the hybrid $\alpha$-attractor model, assuming the {\em Planck} pivot scale $k_{\star}=0.05\,{\rm Mpc}^{-1}$ (indicated by the vertical line) left the horizon 55 $e$-folds before the end of inflation. In all panels the light blue curve corresponds to the solution for the benchmark parameters~\eqref{eq:benchmark}. The remaining spectra show the result of varying $d$ (left panel), $\chi_0$ (center panel) and $g$ (right panel), while keeping the rest of the parameters unchanged (see Table~\ref{tab:hybrid} and the main text for more details).
	\label{fig:PRhybrid}}
\end{figure}

\begin{figure}[t!]
	\centering
	\includegraphics[scale=0.65]{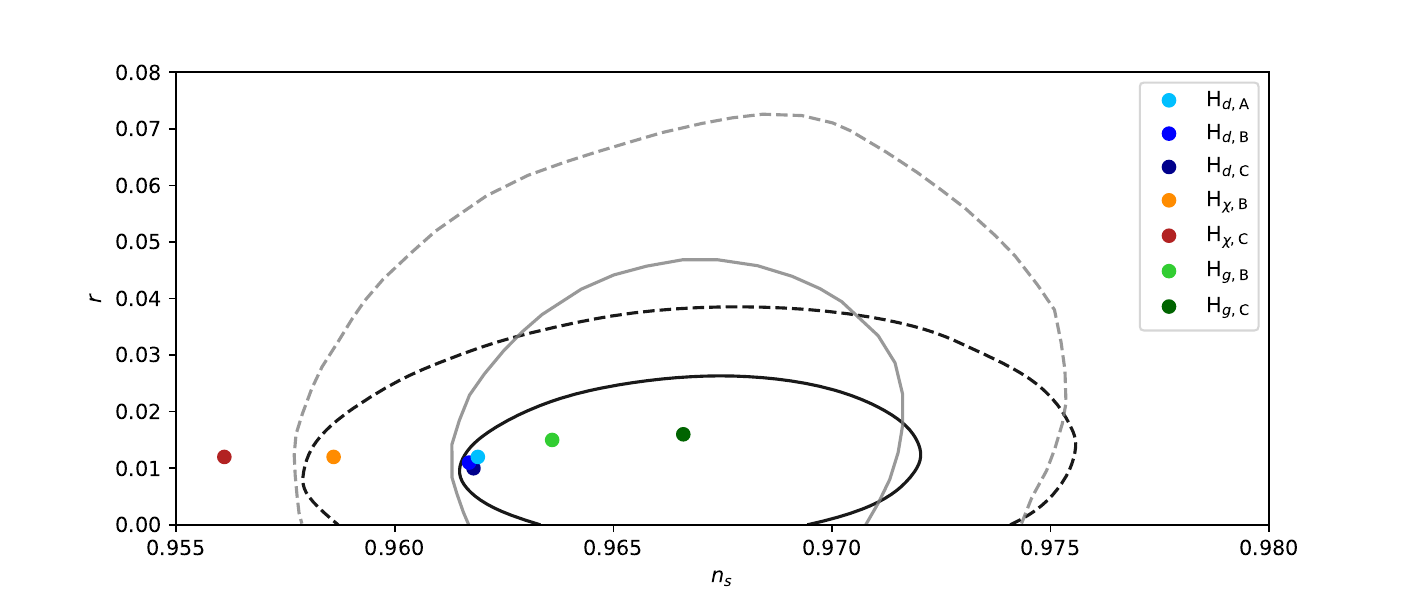}
	\caption{Inflationary CMB observables $n_s$ and $r$ for the hybrid $\alpha$-attractor model. The scalar tilt and the tensor-to-scalar ratio are evaluated under the assumption that the {\em Planck} pivot scale $k_{\star}=0.05\,{\rm Mpc}^{-1}$ left the horizon 55 $e$-folds before the end of inflation. While the light blue point H$_{d,\mathrm{A}}$ is obtained for the benchmark parameters~\eqref{eq:benchmark}, the remaining points result from varying their values, cf.\ Table~\ref{tab:hybrid}.
	Analogously to Fig.~\ref{fig:anisotropies_plot}, the gray and black contour lines arise from the experimental data of \textit{Planck} in combination with BK15+BAO \cite{Planck:2018jri} and BK18+BAO \cite{BICEP:2021xfz}, respectively.
	\label{fig:Hybrid_ns_r}}
\end{figure}

The potential~\eqref{eq:hybridV} is illustrated in Fig.~\ref{fig:HV} for the benchmark parameters~\cite{Braglia:2022phb}
\begin{equation}
\label{eq:benchmark}
M\;=\; 1.47\times 10^{-5}\,,\quad \alpha\;=\;1\,,\quad g=0.8\,,\quad m\;=\;0.3\,,\quad \chi_0\;=\;2.5\,,\quad d \;=\; -5\times 10^{-6}\,.
\end{equation}
The solid blue curve on the surface of the potential corresponds to a consistent inflationary trajectory, which displays
the rapid turn in field space due to the change in sign of the effective mass. Note that after the waterfall transition the motion does not depend only on $\chi$, and therefore multifield methods are appropriate to determine the curvature power spectrum. With the parameters~\eqref{eq:benchmark}, we compute the curvature power spectrum $\mathcal{P}_{\mathcal{R}}$ by solving Eqs.~\eqref{eqs:QsigmaQs}. The result is presented by the light blue curve in all three panels of Fig.~\ref{fig:PRhybrid}. We label this solution as H$_{d,{\rm A}}$ in this figure and also in Table~\ref{tab:hybrid}. The enhancement of the spectrum at small scales (large $k$) is evident, leading to amplitudes as large as $\mathcal{P}_{\mathcal{R}}\sim 10^{-2}$, which may lead to the production of primordial black holes and the emission of an associated background of gravitational waves~\cite{Spanos:2021hpk,Braglia:2022phb}. Importantly, we note that, despite this enhancement, the spectrum is nearly-scale invariant at CMB scales, as the resulting values of $n_s$ and $r$ in Table~\ref{tab:hybrid} and also in Fig.~\ref{fig:Hybrid_ns_r} prove. The large spectrum at scales relevant for the presence of $\mu$-type distortions, $1\,{\rm Mpc}^{-1}\lesssim k \lesssim 10^4\,{\rm Mpc}^{-1}$, suggests a significant enhancement of $\Delta I$ w.r.t.\ the standard $\Lambda$CDM prediction, which must be inspected (see discussion below).

In order to explore the parameter space for this hybrid model, we vary the Lagrangian parameters $d$, $\chi_0$ and $g$ with respect to the benchmark~\eqref{eq:benchmark}, mimicking the analysis in~\cite{Braglia:2022phb}.\footnote{When necessary, the parameter $M$ is adjusted to match the measured amplitude $A_{S^\star}$ at the {\em Planck} pivot scale.} In total we compute the curvature power spectrum for a set of seven points, shown in Table~\ref{tab:hybrid}, labeled by the parameter that is varied with respect to (\ref{eq:benchmark}). The corresponding power spectra are shown in Fig.~\ref{fig:PRhybrid}. In the left panel, the parameter $d$ is varied. We note that upon increasing its (absolute) value, the peak of the distribution decreases, and is shifted to the right. We therefore expect reduced SDs for the darker blue curves. In the middle panel, $\chi_0$ is decreased. In this case, the enhancement of the spectrum remains approximately constant, and the effect is just a shift of the peak. Since this shift moves the blue-tilted region away from the SDs window, we also expect a suppression in the $\Delta I$ signal for the orange and red curves compared to the benchmark spectrum. Finally, the rightmost panel depicts the result of tuning the value of $g$. A larger $g$ results in a larger peak, shifted away from the SDs range. Hence, despite the enhanced power spectrum, the green curves will lead to a smaller distortion imprint. 

Fig.~\ref{fig:Hybrid_ns_r} depicts the numerically evaluated CMB observables for the points shown in Table~\ref{tab:hybrid}, compared with the current observational constraints. We observe that only H$_{\chi,{\rm B}}$ and H$_{\chi,{\rm C}}$, corresponding to an increase in $\chi_0$ relative to the benchmark value~\eqref{eq:benchmark}, lie outside the 1$\sigma$ CL region, and only H$_{\chi,{\rm C}}$ lies outside the 2$\sigma$ CL contours. Interestingly, the benchmark point H$_{d,{\rm A}}$ lies near the edge of the {\em Planck}+BK18 68\% CL contour. This implies that not only this point can be verified or ruled out in the future from its SDs signal, but also from the upcoming improvements in the constraints for $n_s$. 

\begin{figure}[t!]
	\includegraphics[scale=0.26]{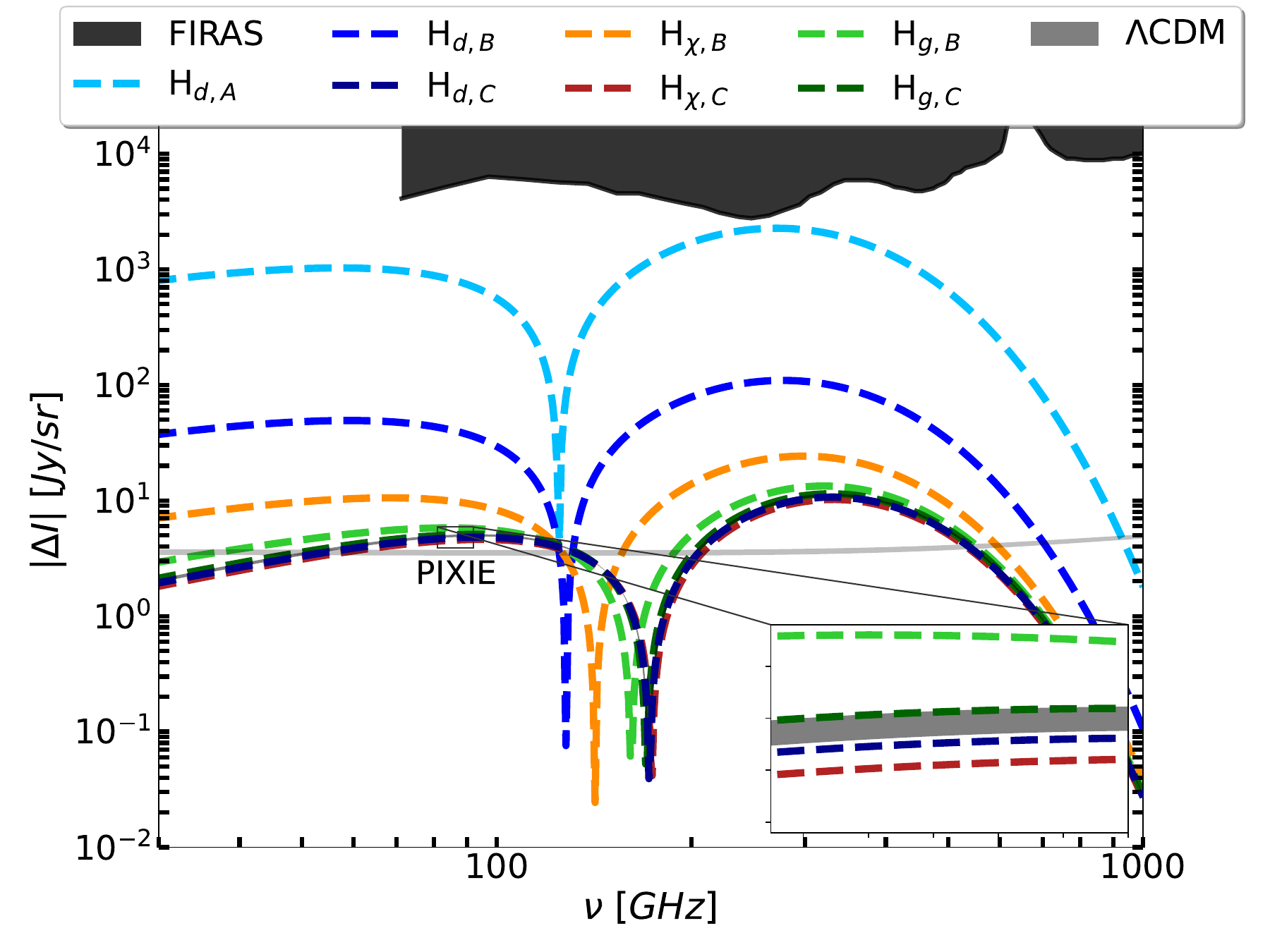}
	\includegraphics[scale=0.26]{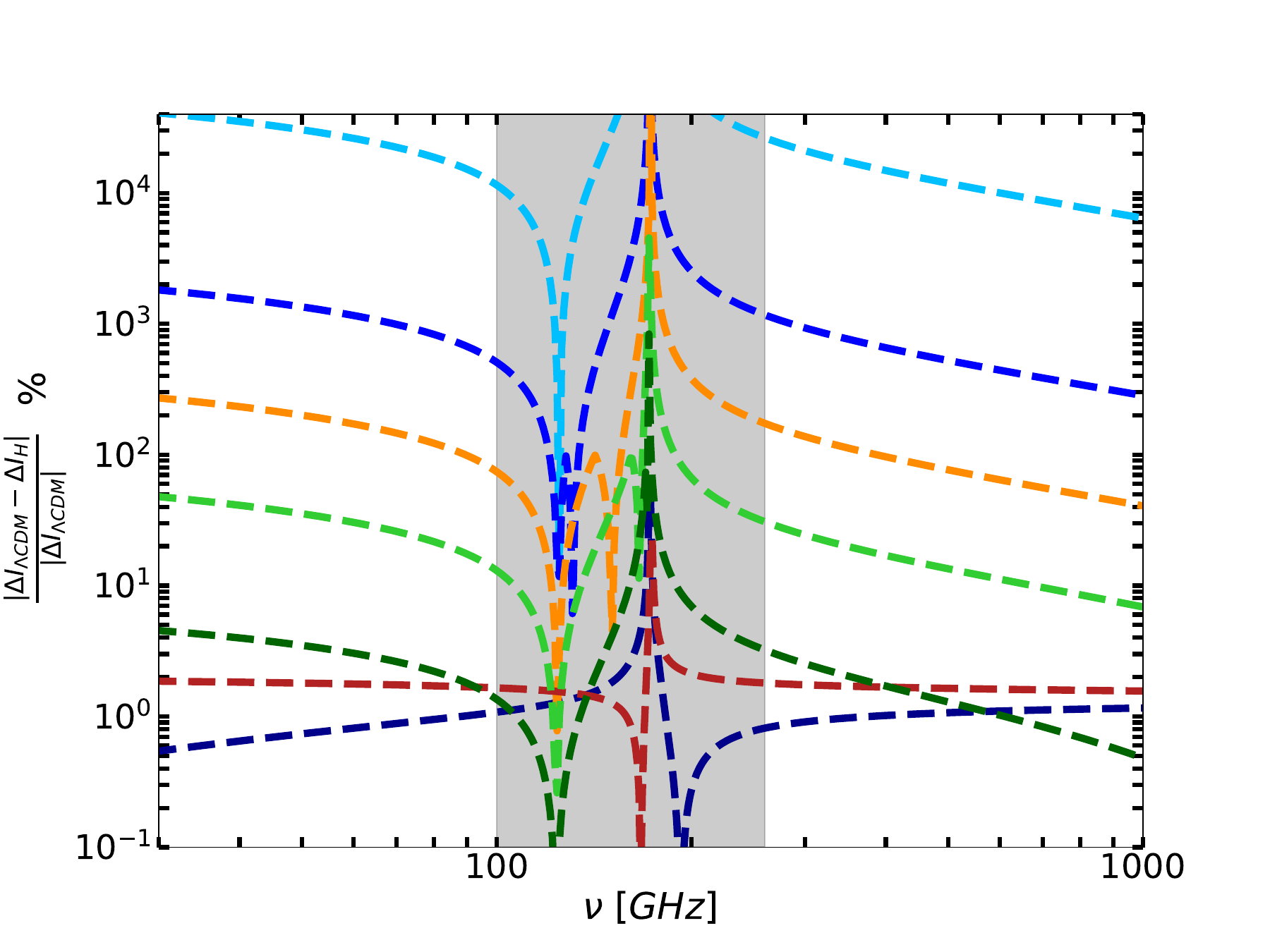}
	\caption{Left: $\Delta I$ prediction for the hybrid inflationary models, contrasted against the $\Lambda$CDM prediction, the FIRAS sensitivity (black region adapted from~\cite[Fig.~1]{Abitbol:2017vwa}) and the PIXIE projected sensitivity. The dashed curves share the color coding with the background solutions and spectra in Fig.~\ref{fig:Hybrid_ns_r}, and are labeled as H$_{d,A}$, H$_{d,B}$, H$_{d,C}$,  H$_{\chi,B}$, H$_{\chi,C}$, H$_{g,B}$ and H$_{g,C}$. Right: percentage difference between the hybrid inflationary prediction $\Delta I_\mathrm{H}$ and $\Delta I_{\Lambda\mathrm{CDM}}$. For frequencies outside the range $100\,\mathrm{GHz}\lesssim\nu\lesssim250\,\mathrm{GHz}$ the difference w.r.t.\ the fiducial signal can be as large as 4$\times$10$^{4}$\%.
        \label{fig:sdhybrid}}
\end{figure}

The resulting SDs for the seven points listed in Table~\ref{tab:hybrid} are shown as dashed curves in the left panel of Fig.~\ref{fig:sdhybrid}, alongside the $\Lambda$CDM prediction and the projected sensitivity of PIXIE. We remark that the distortion $\Delta I_\mathrm{H}$ from a hybrid $\alpha$-attractor model can significantly deviate from the $\Lambda$CDM signal, as to be expected from the form of the power spectrum displayed in Fig.~\ref{fig:PRhybrid}.
\enlargethispage{\baselineskip}
For the discussed sample points we observe that $\Delta I_\mathrm{H}$ deviates from $\Delta I_{\Lambda\mathrm{CDM}}$ as little as 0.5\%, just as in the single-field scenario, or as much as 4$\times$10$^{4}$\% in the relevant region $\nu\lesssim100$\,GHz and $\nu\gtrsim250$\,GHz. This is more explicitly conveyed in the right panel of Fig.~\ref{fig:sdhybrid}.

\section{Summary of results}
\label{sec:results}
We have studied a selection of promising single-field and multifield inflationary models in search for predictions for detectable signals of SDs by forthcoming missions that aim at a better resolution of the absolute intensity of the CMB, such as PIXIE. We have surveyed their parameter space, aiming at a simultaneous compatibility with current measurements of the most relevant CMB inflationary observables: $A_{S_\star}, n_s, r$ and ${\mathrm dn_s}/{\mathrm d\ln k}$. In particular, we have studied the set of  single-field inflationary scenarios known as axion monodromy, multi-natural, spontaneously broken SUSY, (quadratic and quartic) hilltop inflation, and the T-model. In these cases, after numerically computing their scalar power spectrum, we have performed a parametrical $\chi^2$ analysis in order to identify viable benchmark configurations that  
fit the observables, and then computed the emerging SDs in each case by using the numerical \texttt{CLASS} code. In addition, as a sample of multifield inflationary models, we have considered two-field cases of quadratic multifield inflation, the EGNO model and hybrid attractors. After discussing and computing their more complex structure of power spectra, we have identified benchmark scenarios that comply with inflationary bounds and computed numerically their prediction for the amplitude of the $\Delta I$ signal associated with SDs.

In order to compare the predictions for SDs of the chosen inflationary models, we select benchmark
parameter configurations for each model and study here the resulting heating rate and the amplitude of the predicted signal of SDs. 

\begin{figure}[t!]
	\centering
	\includegraphics[scale=0.26]{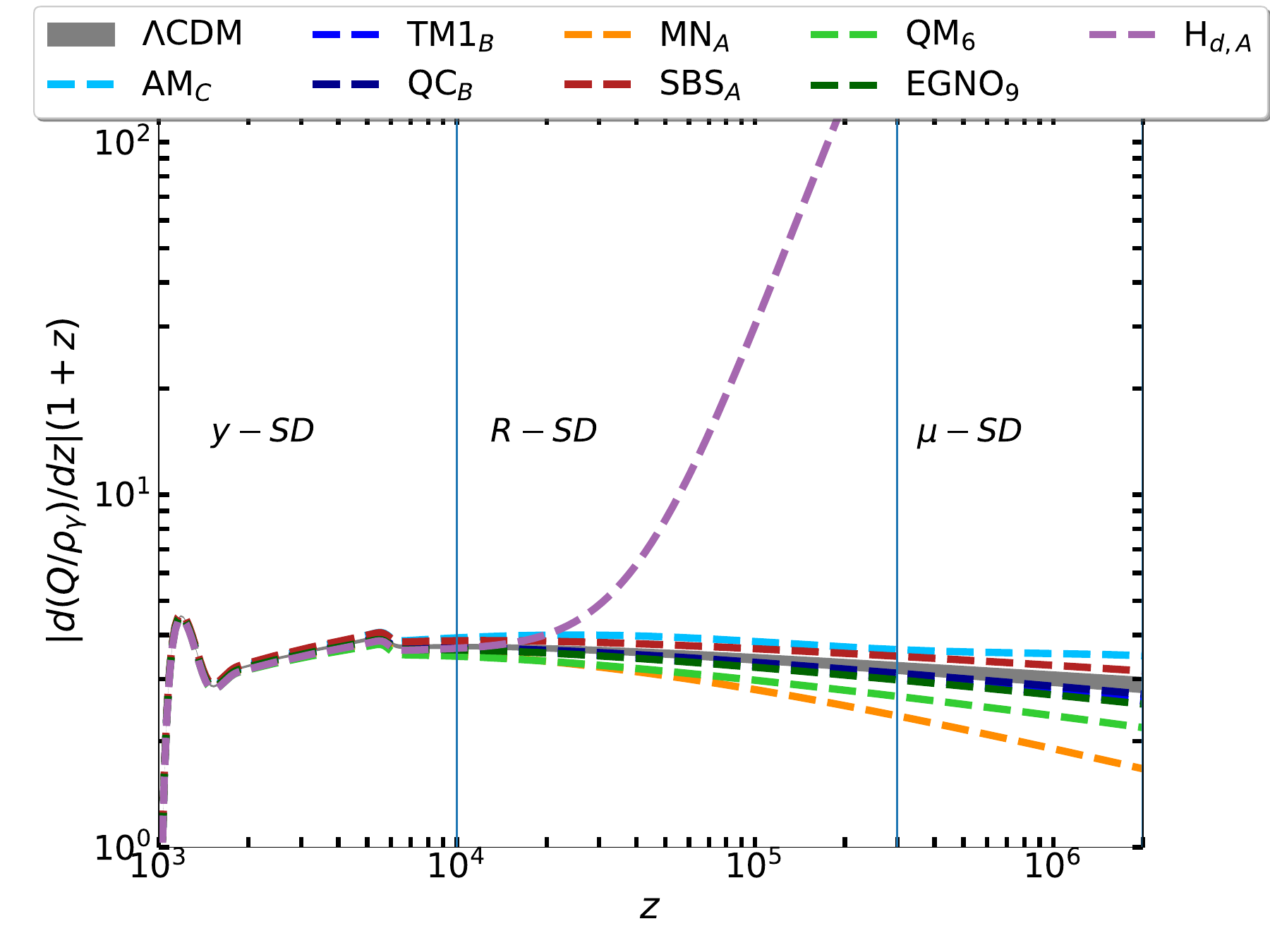}
	\caption{Heating rate $\rho_{\gamma}^{-1}\mathrm d Q/\mathrm dz$ corresponding to the benchmark
		configurations of a selection of inflationary models. We display the predictions for the cases AM$_{C}$ (axion monodromy), TM1$_{B}$  (T-model with $n=1$), QC$_{B}$ (quartic hilltop), MN$_{A}$ (multi-natural inflation), SBS$_{A}$ (spontaneously broken SUSY), QM$_{6}$ (quadratic multifield inflation),  EGNO$_{9}$ and H$_{d,A}$ (Hybrid attractors), together with the $\Lambda$CDM prediction for comparison. See Sections~\ref{sec:SingleFieldInflation} and~\ref{sec:multifield} for details of the cases.
	\label{fig:dqdz}}
\end{figure}

Fig.~\ref{fig:dqdz} allows for a comparison of the various contributions to the heating rate arising from the standard $\mu$, $R$ and $y$ stages of SDs. We confirm that, as already pointed out in Sections~\ref{sec:multinatural} and~\ref{sec:axion-monodromy}, multi-natural and axion-monodromy inflationary models deliver the largest heating rates among the single-field models. Meanwhile, the most significant contribution of all scenarios arises from the two-field inflationary model based on hybrid attractors, as discussed in Section~\ref{sec:hybrid}. As expected the heating rate receives the largest contributions during the $\mu$ stage; however, note that there are still non-fully-negligible contributions to the heating rate at $z\lesssim10^4$ that differ from the standard $\Lambda$CDM expectation coming from most of the distinguishable frameworks that we have studied. Curiously enough, hybrid attractors do not seem to differ from $\Lambda$CDM for smaller red-shifts, stressing already its compatibility with observations at small $z$.

\begin{figure}[t!]
    \centering
    \includegraphics[scale=0.26]{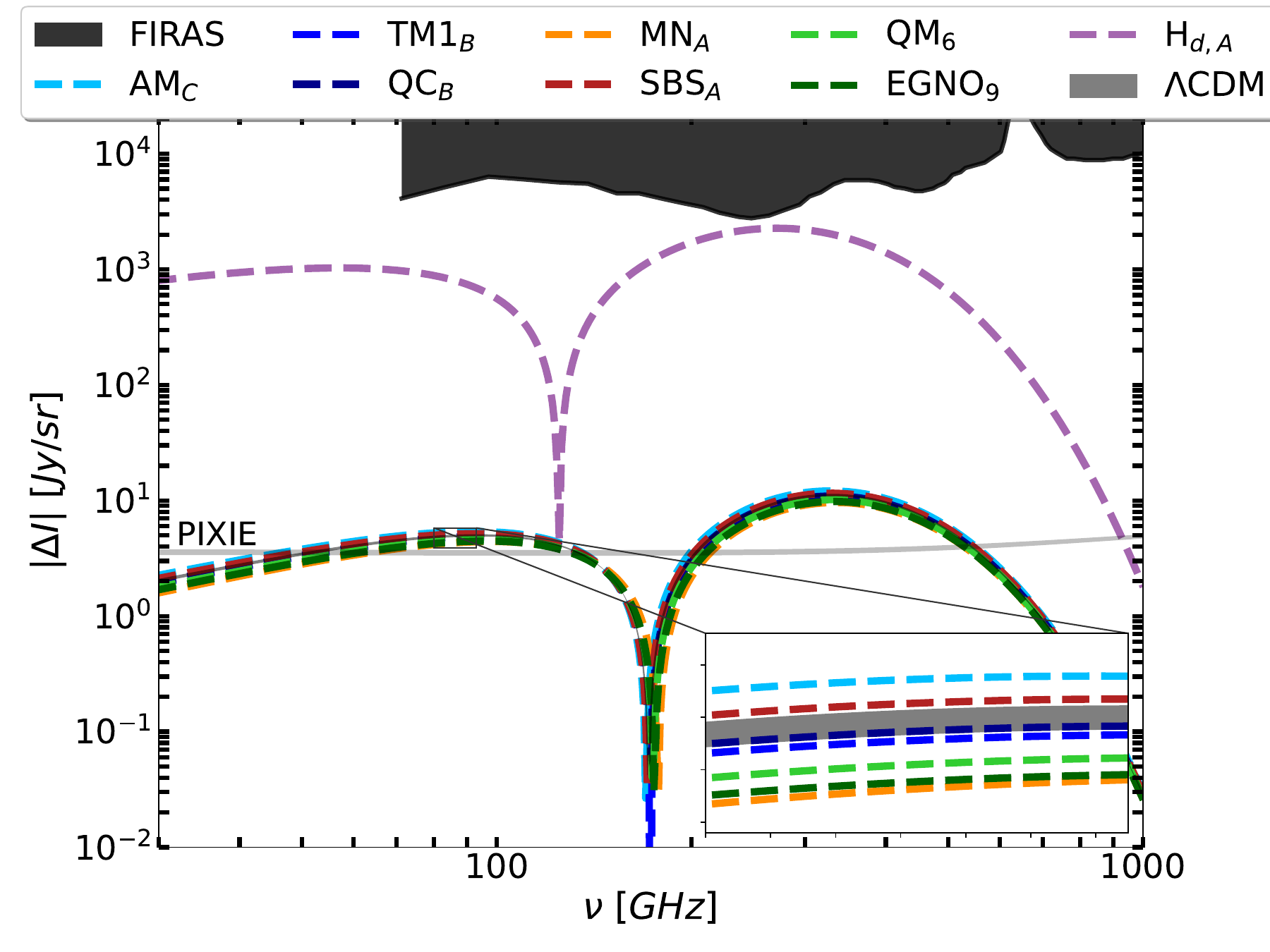}
    \includegraphics[scale=0.26]{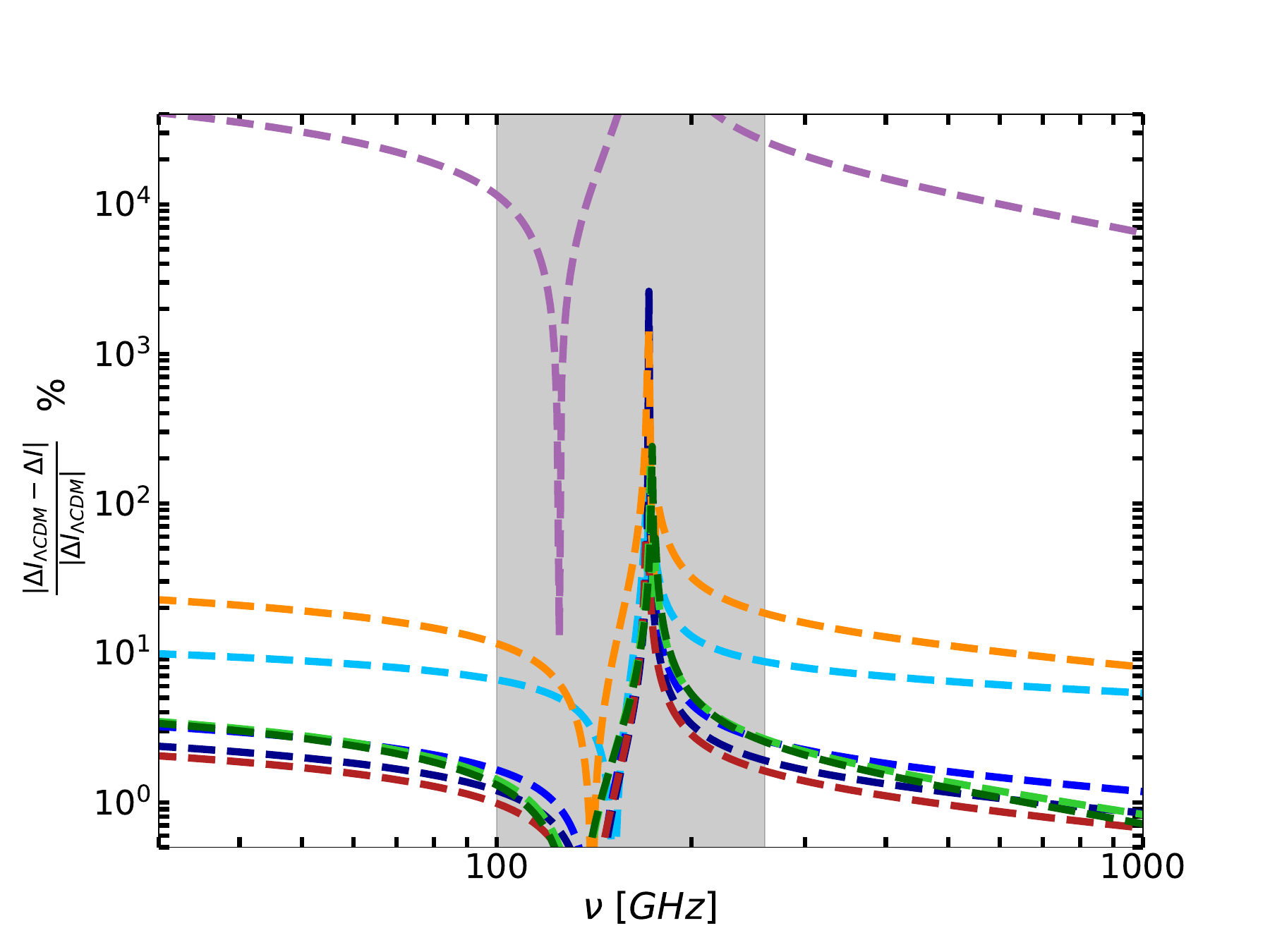}	
    \caption{Left: predicted $\Delta I$ of the photon intensity~\eqref{eq:sdn2} for the benchmark
    	scenarios of a selection of promising single-field and multifield inflationary models, contrasted the $\Lambda$CDM prediction and the forecast of the sensitivity of PIXIE~\cite{Chluba:2019nxa}. Right: for better visibility, we express the difference between the predictions of our models and that of the standard $\Lambda$CDM model as a percentage of the latter. The inflationary model based on hybrid attractors discussed in Section~\ref{sec:hybrid} offers the best scenario of detectability of SDs.
    \label{fig:sd_plots_all}}
\end{figure}

On the other hand, similar to the individual results on the $\Delta I$ predictions for the models studied in the previous sections, we present in Fig.~\ref{fig:sd_plots_all} a comparison of the emerging signals of SDs arising from the benchmark configurations 
of our models. As expected from the behavior of the heating rate, the single-field scenarios that exhibit a best chance to be confronted with observations are those arising from multi-natural inflation and axion monodromy, both cases based on the unconfirmed existence of axion-like particles. These cases present about 10\%-20\% difference for some momenta $k$ with respect to the expectation of SDs from the standard $\Lambda$CDM model, which will challenge the accuracy of future observations to clearly distinguish these discrepancies from the standard signal.
Interestingly, the multi-field scenarios provide new opportunities. The most physically relevant signal in our scan emerges from a model based on two-field hybrid attractors. In this case, the pattern of SDs describes a difference of up to 100 times that of standard $\Lambda$CDM. This outstanding result represents an interesting phenomenological opportunity for forthcoming probes as this amplitude of SDs could be either i) easily distinguished from the background or ii) discarded rapidly. Furthermore, since this signal is associated with a power spectrum rising to values as large as $\mathcal P_{\mathcal R}\sim 10^{-2}$ at momenta as large as $k\sim10^{10}$\,Mpc$^{-1}$, we expect simultaneously a large production of primordial black holes and the emission of an associated background of gravitational waves (as has been suggested recently~\cite{Cyr:2023pgw,Tagliazucchi:2023dai}). This opens up the question of a possible correlation about these observables and SDs. Note though in Fig.~\ref{fig:PRhybrid} that there are even higher peaks (with correspondingly large production of black holes and gravitational waves) that do not yield large SDs. This puzzle will be explored in detail elsewhere.

\vspace{-5mm}
\subsection*{Acknowledgments}
MG and MHN are supported by the DGAPA-PAPIIT grant IA103123 at UNAM, and the CONAHCYT ``Ciencia de Frontera'' grant CF-2023-I-17.
SR-S was supported by UNAM-PAPIIT grant IN113223, CONACyT grant CB-2017-2018/A1-S-13051, PIIF,  and Marcos Moshinsky Foundation.


{\small

\providecommand{\bysame}{\leavevmode\hbox to3em{\hrulefill}\thinspace}

}
\end{document}